\documentclass[aps,pra,reprint,amsfonts,amsmath,amssymb]{revtex4-1}
\usepackage{graphicx}
\usepackage{dcolumn}
\usepackage{bm}
\usepackage{mathtools}
\usepackage{newtxtext}
\usepackage[varg]{newtxmath}
\usepackage[usenames]{color}
\usepackage[normalem]{ulem}
\usepackage{lipsum}

\DeclareMathOperator{\real}{Re}
\DeclareMathOperator{\imaginary}{Im}
\DeclareMathOperator*{\Tr}{Tr}
\DeclareMathOperator{\sgn}{sgn}

\hypersetup{colorlinks=true,linkcolor=blue,citecolor=blue,urlcolor=blue}
\begin{document}

\title{Photocontrol of magnetic structure in an itinerant magnet}
\author{Atsushi Ono}
\author{Sumio Ishihara}
\affiliation{Department of Physics, Tohoku University, Sendai 980-8578, Japan}

\begin{abstract}
We study the photoinduced magnetic transition in an itinerant magnet described by the double-exchange model, in which conduction electrons couple with localized spins through the ferromagnetic (FM) Hund coupling.
It is shown that intense light applied to the FM ground state induces an antiferromagnetic (AFM) order, in contrast to the AFM-to-FM transition due to the photocarrier injection.
In particular, we focus on the mechanism for instability of the FM structure by the light irradiation.
The magnon spectrum in the Floquet state is formulated on the basis of the pertrubative expansion of the Floquet Green function.
The magnon dispersion shows softening at momentum $(\pi,\pi)$ in the square lattice with increasing the light amplitude, implying photoinduced AFM instability.
This result is mainly attributed to a nonequilibrium electron distribution, which promotes low-energy Stoner excitations.
The transient optical conductivity spectra characterized by interband excitations and Floquet sidepeaks are available to identify the photoinduced AFM state.
\end{abstract}
\maketitle

\section{Introduction}
Ultrafast optical control of magnetism has attracted much interest in the past two decades, accompanied by rapid progress in laser light technologies~\cite{Kirilyuk2010,Mentink2017,Kampfrath2013}.
After the pioneering work on the ultrafast demagnetization due to the rapid spin-temperature increase~\cite{Beaurepaire1996}, various strategies to control magnetism have been proposed and demonstrated.
Among them, photoinduced phase transisions involved with magnetic phase transitions make it possible to control magnetism in picosecond or femtosecond timescales owing to the multiple degrees of freedom of electrons and strong correlation between them~\cite{Nasu2004,Tokura2006,Basov2011}.
Another approach called the Floquet engineering is known as an efficient technique to control the electron-electron interaction directly and non-thermally using a time-periodic field~\cite{Mentink2017,Eckardt2017,Oka2018}.
Many proposals for novel Floquet states have been made along this direction~\cite{Takayoshi2014b,Takayoshi2014,Itin2015,Mentink2015,Mikhaylovskiy2015,Sato2016,Bukov2016,Eckstein2017,Kitamura2017,Takasan2017a,Duan2018,Gorg2018,Liu2018,Barbeau2018}.

One of the prototypical ferromagnetic (FM) interactions in metals is the double-exchange (DE) interaction.
This was originally proposed by Zener and Anderson--Hasegawa for FM oxides in 1950s~\cite{Zener1951,Anderson1955,DeGennes1960}.
An essential element of the DE interaction is a strong intra-atomic exchange interaction between mobile electrons and localized spins, which favors the FM configuration.
Therefore, the electronic transport and the magnetism strongly correlate with each other in the DE systems.
This correlation has been ubiquitously observed not only in the FM oxides but also in magnetic semiconductors~\cite{Ohno1999,Hellman2017}, $f$-electron systems~\cite{Yanase1968}, and molecular magnets~\cite{Bechlars2010}, and has described a number of phenomena such as the colossal magnetoresistance~\cite{Tokura1996,Dagotto2001}, the anomalous Hall effect~\cite{Ye1999,Tatara2002,Nagaosa2010,Weng2015}, and skyrmion physics~\cite{Nagaosa2013,Ozawa2017}.

Photoinduced dynamics in the DE system has also been investigated experimentally~\cite{Kiryukhin1997,Miyano1997,Koshihara1997,Fiebig1998,Averitt2001,Rini2007,Matsubara2007,Ichikawa2011,Zhao2011,Yada2016,Lin2018} and theoretically~\cite{Chovan2006,Matsueda2007,Kanamori2009,Kanamori2010,Ohara2013,Koshibae2009,Koshibae2011}, in particular, in perovskite manganites.
Most of those studies have focused on the photoirradiation effects in insulating phases with an antiferromagnetic (AFM) long-range order, and showed formation of a metallic FM domain or an increase in the FM correlation.
These experimental observations are well interpreted by extension of the DE scenario;
photoinjected carriers mediate the DE interaction even though the system is out of equilibrium.

In this paper, we study the photoinduced nonequilibrium dynamics in the DE model.
In Ref.~\cite{Ono2017}, the authors have numerically demonstrated that an initial FM metal state is changed to an almost perfect N\'eel state by photoirradiation, which is in sharp contrast to the naive DE scenario in equilibrium states.
In order to elucidate the microscopic mechanism that drives the FM state into the AFM state, here we study the magnetic structure in a continuous-wave (cw) field by using the Floquet Green function.
We show that a magnon dispersion is softened and has a dip at momentum $\bm{q}=(\pi,\pi)$ by the photoirradiation, which indicates that the AFM instability develops at finite threshold intensity.
It is revealed that a nonequilibrium electron distribution plays an essential role to induce the instability.
We also calculate the transient optical conductivity spectra in a nonequilibrium state through the real-time simulation, and show that an interband-excitation peak and Floquet sidepeaks appear in the transient and steady states.

This paper is organized as follows.
We describe our formulation including a model Hamiltonian and numerical methods in Sec.~\ref{sec:formulation}.
Section~\ref{sec:results} consists of two parts:
first we show the results of the real-time dynamics in Sec.~\ref{sec:real-time}, and then show the magnetic excitation spectra in the photoirradiated FM metal by using the Floquet Green function in Sec.~\ref{sec:spectra}.
Section~\ref{sec:summary} is devoted to a summary.

\section{Formulation} \label{sec:formulation}
First, we introduce the model Hamiltonian and the Floquet Green function method in Secs.~\ref{sec:model} and \ref{sec:selfenergy}.
Next, we derive expressions of the optical conductivity in Sec.~\ref{sec:conductivity}, which is used in the real-time simulation given in Sec.~\ref{sec:time-evol}.

\subsection{Model} \label{sec:model}
We adopt the DE model defined by the Hamiltonian:
\begin{align}
H = \sum_{ij s} h_{ij} c_{is}^\dagger c_{js} - \frac{J}{S} \sum_{iss'} \bm{S}_i \cdot \bm{\sigma}_{ss'} c_{is}^\dagger c_{is'},
\label{eq:hamiltonian}
\end{align}
where $c_{is}^\dagger (c_{is})$ is a creation (annihilation) operator of a conduction electron with spin $s \ (={\uparrow},{\downarrow})$ at site $i$, $\bm{S}_i$ is a localized-spin operator with magnitude $S$, and $\sigma^\alpha \ (\alpha=x,y,z)$ are the Pauli matrices.
The first term ($H_0$) represents the hopping of the conduction electrons with the transfer integral $h_{ij}$, and the second term ($V$) represents the Hund coupling between the conduction electrons and the localized spins with the coupling constant $J\ (>0)$.
The total number of sites and that of electrons, and the electron density are denoted by $N$, $N_e$, and $n_e\equiv N_e/N$, respectively.
Static and dynamical properties in equilibrium states in the DE model have been intensively studied to date, and the FM metallic phase is realized in a wide parameter range $J\gtrsim 4$ and $n_e\lesssim 0.8$~\cite{Yunoki1998}.
A vector potential $\bm{A}$ of light is introduced as the Peierls phase as $h_{ij} \mapsto h_{ij} \exp[ ie\bm{A}({t})(\bm{r}_i-\bm{r}_j) /\hbar ]$, where ${t}$ represents time, $\bm{r}_i$ is a position vector at site $i$, and $e\ (<0)$ is the electron charge.
We adopt the cw field for which the vector potential is given by
$
\bm{A}({t}) = (\bm{F}_0/\mathit{\Omega}) \sin(\mathit{\Omega}{t}),
$
where $\bm{F}_0$ and $\mathit{\Omega} \equiv 2\pi/T$ are amplitude and frequency of the electric field, respectively.
The calculations for a pulse field are presented in Ref.~\cite{Ono2017}.
We consider the two-dimensional square lattice with the lattice constant $a$.
The transfer integral $h_{ij}$ is given by $h_{ij} = -h\ (<0)$ in the nearest-neighbor bonds and $h_{ij}=0$ in the others.
Energy and time are measured in units of $h$ and $\hbar/h$, respectively.
From now on, the nearest-neighbor hopping amplitude $h$, the reduced Planck constant $\hbar$, the electron charge $e$, and the lattice constant $a$ are taken to be unity.

In order to carry out a perturbative expansion which will be introduced in Sec.~\ref{sec:selfenergy}, we rewrite the Hamiltonian by the Holstein--Primakoff transformation for the localized spin operators as
\begin{align}
S_i^z &= S - a_i^\dagger a_i, \quad
S_i^+ = (S_i^-)^\dagger = \sqrt{2S} \sqrt{1-\frac{a_i^\dagger a_i}{2S}} a_i ,
\label{eq:holstein-primakoff}
\end{align}
where $a_{i}^\dagger\ (a_{i})$ is a creation (annihilation) operator of a magnon at site $i$.
Up to the leading order in $1/S$, the Hund coupling term of the Hamiltonian is written as
\begin{align}
V &= -\frac{J}{S} \sum_i \Bigl[ (c_{i\uparrow}^\dagger c_{i\uparrow} - c_{i\downarrow}^\dagger c_{i\downarrow})(S-a_i^\dagger a_i) \notag \\
&\quad + \sqrt{2S} (c_{i\uparrow}^\dagger c_{i\downarrow} a_i^\dagger + c_{i\downarrow}^\dagger c_{i\uparrow} a_i) \Bigr].
\label{eq:hundterm}
\end{align}
By introducing the Fourier transformations for the electron and magnon operators,
\begin{align}
c_{\bm{k}s} = \frac{1}{\sqrt{N}} \sum_{i} e^{-i\bm{k}\bm{r}_i} c_{is}, \quad
a_{\bm{k}} = \frac{1}{\sqrt{N}} \sum_{i} e^{-i\bm{k}\bm{r}_i} a_{i},
\end{align}
we redefine $H_0$ as
\begin{align}
H_0 &= \sum_{\bm{k}s} \varepsilon_{\bm{k}s} c_{\bm{k}s}^\dagger c_{\bm{k}s} + \sum_{\bm{q}} \omega_{\bm{q}}^{(0)} a_{\bm{q}}^\dagger a_{\bm{q}},
\label{eq:hamiltonian_h0}
\end{align}
where the second term with $\omega_{\bm{q}}^{(0)}=0$ is added for convenience in the formalism.
The electron band $\varepsilon_{\bm{k}s}$ is defined by $\varepsilon_{\bm{k}s} = \varepsilon_{\bm{k}}-J \sgn(s)-\mu$ including the chemical potential $\mu$, where $\varepsilon_{\bm{k}}$ and $\sgn(s)$ are given by
\begin{align}
\varepsilon_{\bm{k}} = -2(\cos k_x + \cos k_y),
\label{eq:band}
\end{align}
and
\begin{align}
\sgn(s={\uparrow})=+1,\quad \sgn(s={\downarrow})=-1,
\end{align}
respectively.
Equation~\eqref{eq:hundterm} is also rewritten as
\begin{align}
V &= \frac{J}{SN} \sum_{\bm{k}\bm{k}'\bm{q}\bm{q}'s} \delta_{\bm{k}+\bm{q},\bm{k}'+\bm{q}'} \sgn(s) c_{\bm{k}s}^\dagger c_{\bm{k}'s} a_{\bm{q}}^\dagger a_{\bm{q}'} \notag \\
&\quad - J \sqrt{\frac{2}{SN}} \sum_{\bm{k}\bm{q}} {\left( c_{\bm{k}\uparrow}^\dagger c_{\bm{k}+\bm{q}\downarrow} a_{\bm{q}}^\dagger + c_{\bm{k}\downarrow}^\dagger c_{\bm{k}-\bm{q}\uparrow} a_{\bm{q}} \right)},
\label{eq:hamiltonian_v}
\end{align}
where the first and second terms, respectively, termed $V_1$ and $V_{2}$, originate from the longitudinal term ($S^z \sigma^z$) and the transverse terms ($S^x \sigma^x + S^y \sigma^y$) in the Hund coupling.

The system before light irradiation is assumed to be a fully polarized FM state in which all of the conduction-electron spins and the localized spins are directed along the $+z$ direction.
The initial state wave function is given by
\begin{align}
\vert \Psi_0 \rangle = \prod_{\bm{k}}^{\mathclap{\varepsilon_{\bm{k}\uparrow}<0}} c_{\bm{k}\uparrow}^\dagger \vert 0 \rangle,
\label{eq:initialstate}
\end{align}
where $\vert 0 \rangle$ is a vacuum for the electrons and magnons.

\subsection{Floquet Green function and selfenergy} \label{sec:selfenergy}
In this section we introduce the Floquet Green function and derive a magnon selfenergy using the perturbative expansion with respect to the Hund coupling.
This method is based on the Keldysh formalism, which is briefly summarized in Appendices~\ref{sec:keldysh} and \ref{sec:floquet-green}.

We define the full and bare contour-ordered Green functions for the electrons as
%\begin{subequations}
\begin{align}
G_{\bm{k}s}({t},{t}')
&= -i \langle \mathcal{T}_{\mathcal{C}} c_{\bm{k}s}({t}) c_{\bm{k}s}^\dagger({t}') \rangle, \\
\mathcal{G}_{\bm{k}s}({t},{t}')
&= -i \langle \mathcal{T}_{\mathcal{C}} c_{\mathrm{I}\bm{k}s}({t}) c_{\mathrm{I}\bm{k}s}^\dagger({t}') \rangle ,
\end{align}
and those for the magnons as
\begin{align}
D_{\bm{q}}({t},{t}')
&= -i \langle \mathcal{T}_{\mathcal{C}} a_{\bm{q}}({t}) a_{\bm{q}}^\dagger({t}') \rangle, \\
\mathcal{D}_{\bm{q}}({t},{t}')
&= -i \langle \mathcal{T}_{\mathcal{C}} a_{\mathrm{I}\bm{q}}({t}) a_{\mathrm{I}\bm{q}}^\dagger({t}') \rangle ,
\end{align}%\end{subequations}
respectively, where $\langle {\cdots} \rangle \equiv \langle \Psi_0 \vert \cdots \vert \Psi_0 \rangle$ represents the expectation value with respect to the initial state in Eq.~\eqref{eq:initialstate}, and the subscript `$\mathrm{I}$' means the interaction picture (see Appendix~\ref{sec:keldysh}).
As mentioned in Sec.~\ref{sec:model}, the time-dependent vector potential $\bm{A}(t)$ is introduced in the transfer integral as the Peierls phase, which causes a momentum shift in the energy band: $\varepsilon_{\bm{k}} \mapsto \varepsilon_{\bm{k}-\bm{A}(t)}$.
In the Floquet representation introduced in Appendix~\ref{sec:floquet-green}, the inverses of the bare Green functions including a bath selfenergy are given by
%\begin{subequations}\label{eq:green_bare_inv}
\begin{align}
&(\mathcal{G}^{\mathrm{R},-1}_{\bm{k}s})_{mn}(\omega) \notag \\
&= \delta_{mn} (\omega+n\mathit{\Omega}+J\sgn(s)+\mu+i\mathit{\Gamma}) - \bar{\varepsilon}_{mn,\bm{k}}, \label{eq:green_retarded_inv} \\
&(\mathcal{G}^{\mathrm{K},-1}_{\bm{k}s})_{mn}(\omega)
= 2i\delta_{mn} ( 1-2f(\omega+n\mathit{\Omega}) ) \mathit{\Gamma}, \label{eq:green_keldysh_inv}
\end{align}%\end{subequations}
for the retarded and Keldysh Green functions of the electrons, respectively.
Here we introduce coupling strength between the system and the bath, $\mathit{\Gamma}\ (>0)$, and the Fermi--Dirac distribution function given by $f(\omega) =1/(e^{\beta \omega}+1)$ with inverse temperature $\beta$.
We also define $\bar{\varepsilon}_{mn,\bm{k}} = T^{-1}\int_0^T dt\, e^{i(m-n)\mathit{\Omega} t} \varepsilon_{\bm{k}-\bm{A}(t)}$, which is explicitly written as
%\begin{subequations} \label{eq:floquetband}
\begin{align}
\bar{\varepsilon}_{mn,\bm{k}} = -2\mathcal{J}_{m-n}(F_0/\mathit{\Omega})(\cos k_x+\cos k_y)
\label{eq:floquetband_even}
\end{align}
for $m-n = 0 \bmod 2$, and
\begin{align}
\bar{\varepsilon}_{mn,\bm{k}} = -2i\mathcal{J}_{m-n}(F_0/\mathit{\Omega})(\sin k_x+\sin k_y)
\label{eq:floquetband_odd}
\end{align}%\end{subequations}
for $m-n = 1 \bmod 2$, assuming linearly polarized light along the diagonal direction in the square lattice as $\bm{F}_0=(F_0,F_0)$.
The function $\mathcal{J}_n$ is the $n$th-order Bessel function of the first kind.

%========================================
\begin{figure}[t]\centering
\includegraphics[width=0.8\columnwidth]{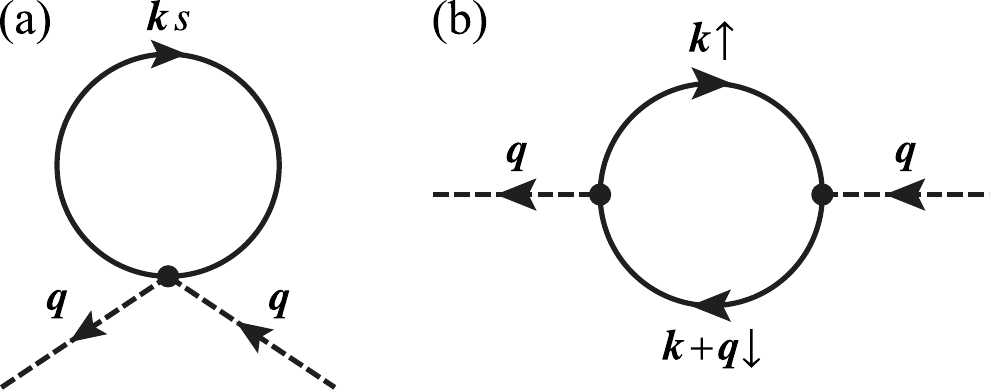}
\caption{The selfenergy diagrams for (a) the longitudinal component $\mathit{\Sigma}_{1}(t,t')$ in Eq.~\eqref{eq:cont_selfenergy1} and (b) the transverse component $\mathit{\Sigma}_{2,\bm{q}}(t,t')$ in Eq.~\eqref{eq:cont_selfenergy2}. Solid and dashed lines represent the free propagators of the electron and magnon, respectively.}
\label{fig:diagram}
\end{figure}
%========================================

First, we consider the longitudinal component of the Hund coupling, $V_1$, and derive the contour-ordered selfenergy $\mathit{\Sigma}_1$ from the first-order expansion of the $S$-matrix in Eq.~\eqref{eq:smatrix} as
\begin{align}
\mathit{\Sigma}_{1}({t},{t}') = \frac{-iJ}{SN} \sum_{\bm{k}s} \sgn(s) \mathcal{G}_{\bm{k}s}({t}',{t}) \delta_{\mathcal{C}}({t},{t}'^+) ,
\label{eq:cont_selfenergy1}
\end{align}
where $\delta_{\mathcal{C}}$ is the contour delta function, and $t'^+$ means the time that is infinitesimally later than $t'$ on the contour.
The corresponding diagram is shown in Fig.~\ref{fig:diagram}(a).
We notice that $\mathit{\Sigma}_1$ is instantaneous and independent of the external momentum.
The off-diagonal components, $\mathit{\Sigma}_1^{12}$ and $\mathit{\Sigma}_1^{21}$, vanish because of $\delta_{\mathcal{C}}({t},{t}'^+)$, while the diagonal components are given as follows:
%\begin{subequations}
\begin{align}
\mathit{\Sigma}_1^{11}({t},{t}') &= \frac{-iJ}{SN} \sum_{\bm{k}s} \sgn(s) \mathcal{G}_{\bm{k}s}^{11}({t}',{t}) [+\delta({t}-{t}'^+) ] \notag \\
&= \delta({t}-{t}') \frac{-iJ}{SN} \sum_{\bm{k}s} \sgn(s) \mathcal{G}_{\bm{k}s}^{12}({t},{t}), \\
\mathit{\Sigma}_1^{22}({t},{t}') &= \frac{-iJ}{SN} \sum_{\bm{k}s} \sgn(s) \mathcal{G}_{\bm{k}s}^{22}({t}',{t}) [ -\delta({t}-{t}'^-) ] \notag \\
&= -\mathit{\Sigma}_1^{11}({t},{t}') ,
\end{align}%\end{subequations}
with $t'^\pm=t'\pm 0$.
Thus, the retarded component of $\mathit{\Sigma}_1$ is given by
\begin{align}
\mathit{\Sigma}^{\mathrm{R}}_1({t},{t}')
&= \frac{1}{2} \left( \mathit{\Sigma}_1^{11} - \mathit{\Sigma}_1^{22} + \mathit{\Sigma}_1^{21} - \mathit{\Sigma}_1^{12} \right) \notag \\
&= \delta({t}-{t}') \frac{-i J}{SN} \sum_{\bm{k}s} \sgn(s) \mathcal{G}_{\bm{k}s}^{<}({t},{t}). \label{eq:selfenergy1_twotime_retarded}
\end{align}
The Floquet representation of the retarded selfenergy, which is termed the Floquet selfenergy, is obtained from Eqs.~\eqref{eq:floquet-twotime} and \eqref{eq:twotime-floquet} as
\begin{align}
(\mathit{\Sigma}_1^{\mathrm{R}})_{mn}(\omega) = \frac{-iJ}{SN} \sum_{\bm{k}s} \sgn(s) \int_{-\infty}^{\infty} \frac{d\bar{\omega}}{2\pi}\, (\mathcal{G}_{\bm{k}s}^{<})_{m-n,0}(\bar{\omega}).
\label{eq:selfenergy1_retarded}
\end{align}

As for the transverse component of the Hund coupling, $V_2$, the contour-ordered selfenergy $\mathit{\Sigma}_2$ is obtained from the second-order term in the $S$-matrix as
\begin{align}
\mathit{\Sigma}_{2,\bm{q}}({t},{t}') = - \frac{2iJ^2}{SN} \sum_{\bm{k}} \mathcal{G}_{\bm{k}+\bm{q}\downarrow}({t},{t}') \mathcal{G}_{\bm{k}\uparrow}({t}',{t}) .
\label{eq:cont_selfenergy2}
\end{align}
The corresponding diagram is shown in Fig.~\ref{fig:diagram}(b).
In a similar way to Eq.~\eqref{eq:selfenergy1_twotime_retarded}, the retarded selfenergy is given by
\begin{widetext}\begin{align}
\mathit{\Sigma}_{2,\bm{q}}^{\mathrm{R}}({t},{t}')
= - \frac{iJ^2}{SN} \sum_{\bm{k}} {\left[ \mathcal{G}_{\bm{k}+\bm{q}\downarrow}^{\mathrm{R}}({t},{t}') \mathcal{G}_{\bm{k}\uparrow}^{\mathrm{K}}({t}',{t}) + \mathcal{G}_{\bm{k}+\bm{q}\downarrow}^{\mathrm{K}}({t},{t}') \mathcal{G}_{\bm{k}\uparrow}^{\mathrm{A}}({t}',{t}) \right]},
\label{eq:selfenergy2_twotime_retarded}
\end{align}
and the corresponding Floquet selfenergy is obtained as
\begin{align}
(\mathit{\Sigma}_{2,\bm{q}}^{\mathrm{R}})_{mn}(\omega)
= - \frac{iJ^2}{SN} \sum_{\bm{k}} \sum_l \int_{-\infty}^{\infty} \frac{d\bar{\omega}}{2\pi} {\left[ (\mathcal{G}_{\bm{k}+\bm{q}\downarrow}^{\mathrm{R}})_{m,n+l}(\omega+\bar{\omega}) (\mathcal{G}_{\bm{k}\uparrow}^{\mathrm{K}})_{l,0}(\bar{\omega}) + (\mathcal{G}_{\bm{k}+\bm{q}\downarrow}^{\mathrm{K}})_{m,n+l}(\omega+\bar{\omega}) (\mathcal{G}_{\bm{k}\uparrow}^{\mathrm{A}})_{l,0}(\bar{\omega}) \right]}.
\label{eq:selfenergy2_retarded}
\end{align}
\end{widetext}
Finally, the total Floquet selfenergy of the magnon is obtained as
\begin{align}
(\mathit{\Sigma}_{\bm{q}}^{\mathrm{R}})_{mn}(\omega)
= (\mathit{\Sigma}_{1}^{\mathrm{R}})_{mn}(\omega) + (\mathit{\Sigma}_{2,\bm{q}}^{\mathrm{R}})_{mn}(\omega). \label{eq:selfenergy_retarded}
\end{align}
The full magnon Green function $D_{\bm{q}}^{\mathrm{R}}$ is given by the Dyson equation:
\begin{align}
(D^{\mathrm{R},-1}_{\bm{q}})_{mn}(\omega) = (\mathcal{D}^{\mathrm{R},-1}_{\bm{q}})_{mn}(\omega) - (\mathit{\Sigma}^\mathrm{R}_{\bm{q}})_{mn}(\omega),
\label{eq:dyson_magnon}
\end{align}
where the bare magnon Green function is given by
\begin{align}
(\mathcal{D}^{\mathrm{R},-1}_{\bm{q}})_{mn}(\omega)
= \delta_{mn}(\omega+n\mathit{\Omega}+i\mathit{\Gamma} - \omega_{\bm{q}}^{(0)}).
\label{eq:dyson_magnon_bare}
\end{align}
In the numerical calculations, the dimension in the Floquet space is limited to $2N_p+1$, and the Floquet indices run over $\{0,\pm 1, \dots, \pm N_p\}$.

We show that the Floquet selfenergy in Eq.~\eqref{eq:selfenergy_retarded} at $\bm{F}_0=0$ and $\mathit{\Gamma} \rightarrow 0$ coincides to the equilibrium selfenergy given in Ref.~\cite{Furukawa1996}.
The bare Green functions are obtained from Eqs.~\eqref{eq:green_retarded_inv} and \eqref{eq:green_keldysh_inv} as
\begin{align}
(\mathcal{G}_{\bm{k}s}^{\mathrm{R}})_{mn}(\omega) &= \frac{\delta_{mn}}{\omega+n\mathit{\Omega}+i\eta-\varepsilon_{\bm{k}s}} = (\mathcal{G}_{\bm{k}s}^{\mathrm{A}})_{nm}(\omega)^*, \\
(\mathcal{G}_{\bm{k}s}^{\mathrm{K}})_{mn}(\omega) &= -2\pi i \delta_{mn} (1-2f(\omega+n\mathit{\Omega})) \delta(\omega+n\mathit{\Omega}-\varepsilon_{\bm{k}s}), \\
(\mathcal{G}_{\bm{k}s}^{<})_{mn}(\omega) &= 2\pi i \delta_{mn} f(\omega+n\mathit{\Omega}) \delta(\omega+n\mathit{\Omega}-\varepsilon_{\bm{k}s}),
\end{align}
where $\eta$ is a positive infinitesimal.
The retarded selfenergy $\mathit{\Sigma}_1^{\mathrm{R}}$ in Eq.~\eqref{eq:selfenergy1_retarded} is expressed as
\begin{align}
\mathit{\Sigma}_1^{\mathrm{R}}
&= (\mathit{\Sigma}_1^{\mathrm{R}})_{nn}(\omega-n\mathit{\Omega}) \notag \\
&= \frac{-iJ}{SN} \sum_{\bm{k}s} \sgn(s) \int_{-\infty}^{\infty} \frac{d\bar{\omega}}{2\pi}\, 2\pi i f(\bar{\omega}) \delta(\bar{\omega}-\varepsilon_{\bm{k}s}) \notag \\
&= \frac{J}{SN} \sum_{\bm{k}} [f(\varepsilon_{\bm{k}\uparrow})-f(\varepsilon_{\bm{k}\downarrow})].
\end{align}
As for the selfenergy $\mathit{\Sigma}_2^{\mathrm{R}}$ in Eq.~\eqref{eq:selfenergy2_retarded}, the contour integral in the complex $\bar{\omega}$ plane gives the following expression:
\begin{align}
\mathit{\Sigma}_{2,\bm{q}}^{\mathrm{R}}(\omega)
&= (\mathit{\Sigma}_{2,\bm{q}}^{\mathrm{R}})_{nn}(\omega-n\mathit{\Omega}) \notag \\
&= \frac{2J^2}{SN} \sum_{\bm{k}} \frac{f(\varepsilon_{\bm{k}\uparrow})-f(\varepsilon_{\bm{k}+\bm{q}\downarrow})}{\omega-(\varepsilon_{\bm{k}+\bm{q}\downarrow}-\varepsilon_{\bm{k}\uparrow})+2i\eta}.
\label{eq:selfenergy_retarded_equil_2}
\end{align}
Therefore, the retarded selfenergy for the magnons takes the following form,
\begin{align}
\mathit{\Sigma}_{\bm{q}}^{\mathrm{R}}(\omega)
&= \mathit{\Sigma}_{1}^{\mathrm{R}} + \mathit{\Sigma}_{2,\bm{q}}^{\rm R}(\omega) \notag \\
&= \frac{J}{SN} \sum_{\bm{k}} [f(\varepsilon_{\bm{k}\uparrow})-f(\varepsilon_{\bm{k}+\bm{q}\downarrow})] \notag \\
&\quad \times {\left[ 1 + \frac{2J}{\omega-(\varepsilon_{\bm{k}+\bm{q}\downarrow}-\varepsilon_{\bm{k}\uparrow})+2i\eta} \right]},
\label{eq:selfenergy_retarded_equil}
\end{align}
which is in agreement with Eq.~(6) in Ref.~\cite{Furukawa1996}.
We note that $\real \mathit{\Sigma}_{\bm{q}=0}^{\mathrm{R}}(\omega=0) = 0$, which ensures the presence of the gapless mode at $\bm{q}=0$ up to the leading order in $1/S$.

\subsection{Optical conductivity} \label{sec:conductivity}
The optical conductivity is defined by a response of the electric current density $\bm{j}=(j^x,j^y)$ to the electric field $\bm{F}=(F^x,F^y)$ as
%\begin{subequations}
\begin{align}
\langle j^\alpha({t}) \rangle
&= \sum_{\beta} \int_{-\infty}^{\infty} d\bar{{t}}\, \sigma_{\alpha\beta}({t},\bar{{t}}) F^\beta(\bar{{t}}) \\
&= \sum_{\beta} \int_{-\infty}^{\infty} d\bar{{t}}\, \chi_{\alpha\beta}({t},\bar{{t}}) A^\beta(\bar{{t}}) ,
\end{align}%\end{subequations}
where $\chi$ is the current susceptibility, satisfying the relation:
\begin{align}
\sigma_{\alpha\beta}({t},{t}') = - \int_{{t}'}^{\infty} d\bar{{t}}\, \chi_{\alpha\beta}({t},\bar{{t}}).
\label{eq:conductivity_twotime}
\end{align}
The optical conductivity at time $t_a$, $\sigma_{\alpha\beta}(\omega,t_a)$, is obtained from the Fourier transformation of the two-time function $\sigma_{\alpha\beta}({t},{t}')$ in Eq.~\eqref{eq:conductivity_twotime} with respect to $t_r = {t}-{t}'$ (see Eq.~\eqref{eq:wigner-twotime}).

The current density and the coupling Hamiltonian between the vector potential and the electrons are given by
\begin{align}
j^{\alpha}({t}) &= \frac{1}{N} \sum_{\bm{k}s} v_{\bm{k}-\bm{A}({t}),s}^\alpha c_{\bm{k}s}^\dagger c_{\bm{k}s},
\label{eq:conductivity_current} \\
V({t}) &= \sum_{\mathclap{\bm{k}s}} (\varepsilon_{\bm{k}-\bm{A}({t}),s} - \varepsilon_{\bm{k},s}) c_{\bm{k}s}^\dagger c_{\bm{k}s},
\label{eq:conductivity_coupling}
\end{align}
respectively, with $\bm{v}_{\bm{k}s}=\partial_{\bm{k}} \varepsilon_{\bm{k}s}$.
According to the general formalism of the response function presented in Appendix~\ref{sec:responsefunction}, we obtain the diamagnetic and paramagnetic responses as
\begin{align}
\chi_{\alpha\beta}^{\mathrm{dia}}({t},{t}')
&= \delta({t}-{t}') \frac{i}{N} \sum_{\bm{k}s} \frac{\partial^2 \varepsilon_{\bm{k}-\bm{A}(t),s}}{\partial k^\alpha \partial k^\beta} G_{\bm{k}s,\bm{k}s}^{<}({t},{t})
\label{eq:response_dia_result}
\end{align}
and
\begin{align}
\chi_{\alpha\beta}^{\mathrm{pm}}({t},{t}')
= -\frac{2}{N} \imaginary \Tr [v^\alpha({t}) G^{\mathrm{R}}({t},{t}') v^\beta({t}') G^{<}({t}',{t})],
\label{eq:response_pm_result}
\end{align}
respectively, where the trace stands for summations over the momentum and spin variables.
Note that Eqs.~\eqref{eq:response_dia_result} and \eqref{eq:response_pm_result} hold even if the Green function has off-diagonal components in the momentum and spin bases.
Thus, these are straightforward extentions of the expressions in Ref.~\cite{Eckstein2008}.

\subsection{Real-time evolution} \label{sec:time-evol}
In this section, we present the numerical method to calculate the real-time dynamics.
A part of this was introduced in Refs.~\cite{Koshibae2009,Koshibae2011,Ono2017}.
We treat the localized spins as classical vectors, which is justified in the limit of large $S$.
Let us suppose that the localized spin configuration $\{\bm{S}_i\}$ is given at time ${t}$.
Then, the Hamiltonian in Eq.~\eqref{eq:hamiltonian} at time $t$ is diagonalized as $H({t})=\sum_{\nu} \varepsilon_\nu({t}) \phi_\nu^\dagger({t}) \phi_\nu({t})$, where $\phi_\nu^\dagger$ is a creation operator of the electron with the single-particle energy $\varepsilon_\nu$.
The wavefunction of the electrons at time ${t}$ is described as a single Slater determinant given by $\vert \Psi({t}) \rangle = \prod_{\nu=1}^{N_e} \psi_\nu^\dagger({t}) \vert 0 \rangle$.
The creation operator $\psi_\nu^\dagger$ is represented by
\begin{align}
\psi_\nu^\dagger({t}) = \sum_{\mu=1}^{2N} \phi_\mu^\dagger({t}) u_{\mu\nu}({t}),
\end{align}
where the unitary matrix $u_{\mu\nu}({t}) = \langle 0 \vert \phi_\mu({t}) \psi_\nu^\dagger({t}) \vert 0 \rangle$ satisfies the initial condition $u_{\mu\nu}({t}=0)=\delta_{\mu\nu}$.
Note that both $\phi_\nu(t)$ and $\psi_\nu(t)$ are the time-dependent operators in the Schr\"odinger picture, because the localized spin configuration $\{\bm{S}_i(t)\}$ depends on time.
When we assume that $\{\bm{S}_i({t})\}$ is fixed during a short time interval $[{t},{t}+\delta{t}]$, the unitary matrix $u_{\mu\nu}({t})$ is given recursively by
\begin{align}
u_{\mu\nu}({t}+\delta {t})
&= \sum_{\lambda=1}^{2N} \langle \mu({t}+\delta{t}) \vert \lambda({t})\rangle e^{i\varepsilon_\lambda({t})\delta{t}} u_{\lambda\nu}({t}) ,
\end{align}
where $\vert \lambda({t}) \rangle = \phi_\lambda^\dagger({t}) \vert 0 \rangle$.
The expectation value of a one-body operator $O({t}) = \sum_{\mu\nu} O_{\mu\nu}({t}) \phi_\mu^\dagger({t}) \phi_\nu({t})$ is given by
\begin{align}
\langle O({t}) \rangle
&= \sum_{\lambda=1}^{N_e} \sum_{\mu=1}^{2N} \sum_{\nu=1}^{2N} u_{\mu \lambda}^*({t}) O_{\mu\nu}({t}) u_{\nu\lambda}({t}) .
\end{align}

The dynamics of the localized spins is described by the Landau--Lifshitz--Gilbert (LLG) equation,
\begin{align}
\frac{\partial\bm{S}_i}{\partial t} = \bm{h}_i^{\mathrm{eff}} \times \bm{S}_i + \alpha \bm{S}_i \times \frac{\partial\bm{S}_i}{\partial t},
\label{eq:llg}
\end{align}
where $\bm{h}_i^{\mathrm{eff}}({t}) = -\langle \partial H({t}) /\partial \bm{S}_i \rangle = (J/S) \sum_{ss'} \langle \bm{\sigma}_{ss'}c_{is}^\dagger c_{is'}\rangle$ is the effective field, and $\alpha$ is the damping constant.
The local spin configuration at time ${t}+\delta{t}$ is calculated with the fixed effective field $\bm{h}^{\mathrm{eff}}_i({t})$ for which the LLG equation is solved analytically~\cite{Koshibae2009}.

In order to calculate the optical conductivity, we introduce the retarded and lesser Green functions
in this formalism as follows:
%\begin{subequations}
\begin{align}
G_{\nu}^{\mathrm{R}}(t_r,t_a)
&= -i\theta(t_r) e^{-i\varepsilon_\nu(t_a)t_r}, \\
G_{\nu}^{<}(t_r,t_a)
&= i n_\nu(t_a) e^{-i\varepsilon_\nu(t_a)t_r}.
\end{align}%\end{subequations}
Here, we define $t_r={t}-{t}'$ and $t_a={t}$, rather than $t_a=(t+t')/2$, to reduce the computational cost, and $n_\nu(t_a) = \langle \phi_{\nu}^\dagger(t_a) \phi_\nu(t_a) \rangle$.
The function $\theta(t)$ is the step function.
These Green functions are reduced to the equilibrium ones given in Eqs.~\eqref{eq:free_twotime_retarded} and \eqref{eq:free_twotime_lesser} in the equilibrium state.
The Green function in the momentum space is given by
\begin{align}
G_{\bm{k}s,\bm{k}'s'}^X(t_r,t_a)
\approx \sum_{\nu} \langle \bm{k}s \vert \nu(t_a) \rangle G_\nu^X(t_r,t_a) \langle \nu(t_a) \vert \bm{k}'s' \rangle ,
\label{eq:timeevol_greenfunction}
\end{align}
for $X=\mathrm{R}$ and ${<}$, where $\vert \bm{k}s \rangle = c_{\bm{k}s}^\dagger \vert 0 \rangle$.
We assume that each single-particle level $\varepsilon_\nu$ and its occupation $n_{\nu}$ are independent of $t_r$, for simplicity.
This function $G_{\bm{k}s,\bm{k}'s'}$ has the off-diagonal components in the momentum and spin bases, because of breakings of the translational symmetry in the real space and the rotational symmetry in the spin space.
We obtain the optical conductivity $\sigma_{\alpha\beta}(\omega,t_a)$ by using Eqs.~\eqref{eq:wigner-twotime}, \eqref{eq:conductivity_twotime}, \eqref{eq:response_dia_result}, and \eqref{eq:response_pm_result} with the Green function in Eq.~\eqref{eq:timeevol_greenfunction}.
The computational time scales as $\mathcal{O}(N^3)$, which is much faster in a nonequilibrium or inhomogeneous system described by a bilinear Hamiltonian than that of a direct evaluation of an extended Kubo formula~\cite{Matsueda2007,Kanamori2009,Kanamori2010,Ohara2013}.

\section{Results}
\label{sec:results}

\subsection{Real-time dynamics} \label{sec:real-time}
%========================================
\begin{figure}[t]
\includegraphics[width=\columnwidth]{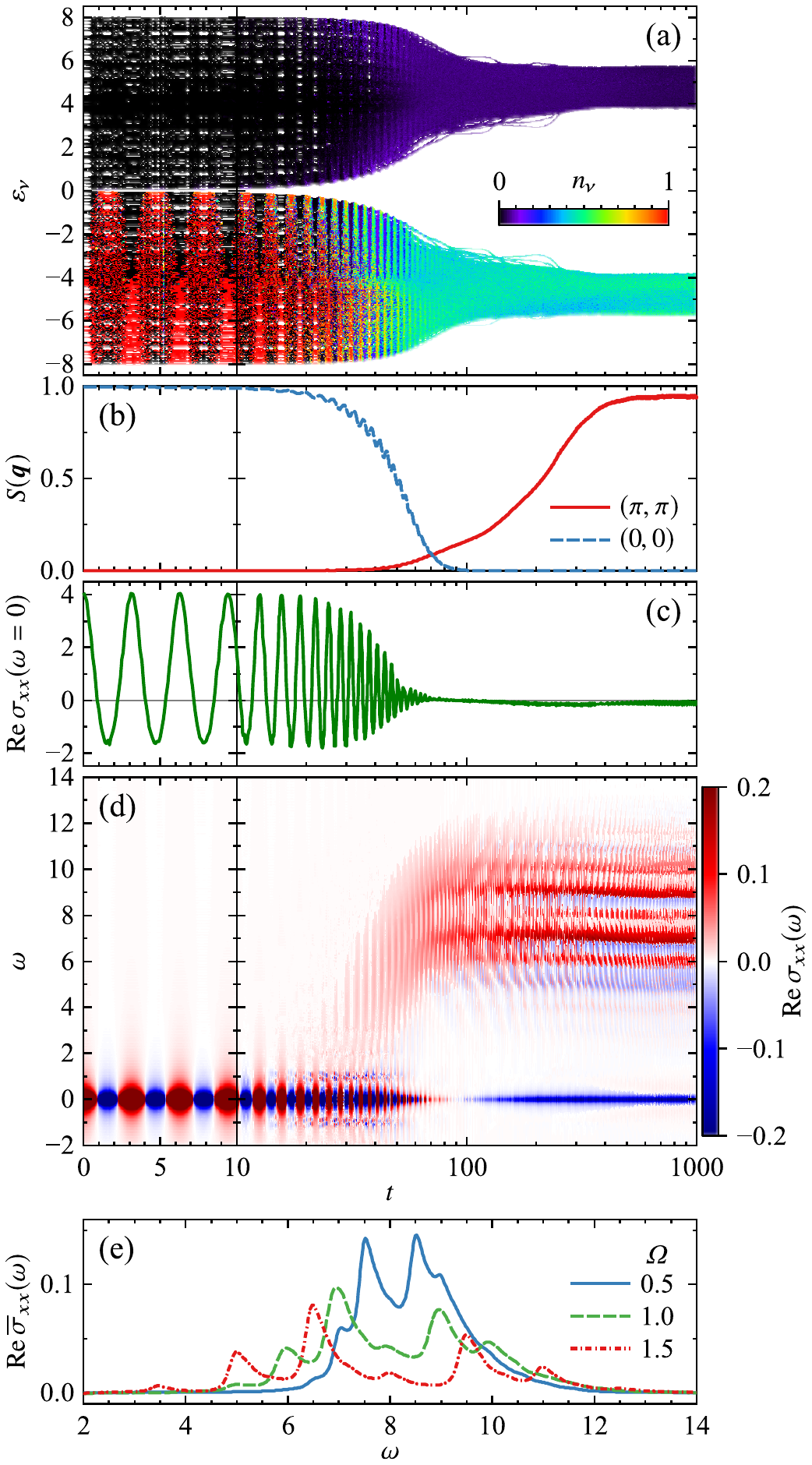}
\caption{(a)--(d) Time profiles of several physical quantities in the real-time dynamics.
(a) The single-particle energy levels $\varepsilon_\nu$ and the occupation numbers $n_{\nu}$.
(b) The spin structure factor $S(\bm{q})$ at $\bm{q}=(0,0)$ and $(\pi,\pi)$.
(c) The static component of the optical conductivity, $\sigma_{xx}(\omega=0,t)$.
(d) The optical conductivity spectrum $\sigma_{xx}(\omega,t)$.
The amplitude and frequency of the cw field are set to $F_0=2$ and $\mathit{\Omega}=1$, respectively.
(e) The time-averaged optical conductivity.
The spectra $\sigma_{xx}(\omega,t)$ for $\mathit{\Omega}=0.5,1.0$, and $1.5$ are averaged during $t=900$ and $1000$.
We chose $F_0/\mathit{\Omega}=2.0$ and $J=4$.}
\label{fig:timeevol}
\end{figure}
%========================================

In this section, we show the real-time dynamics obtained by the method introduced in Sec.~\ref{sec:time-evol}.
We adopt the two-dimensional square lattice with $N=16\times 16$ sites, which is much larger than that in Ref.~\cite{Ono2017}.
The periodic- and antiperiodic-boundary conditions are imposed along the $x$ and $y$ directions, respectively.
The electron number density is set to $n_e=0.5$ (quarter-filling), which provides a FM metallic state in the ground state at $J=4$.
The polarization of the cw field is taken to be the diagonal direction, i.e., $F_0^x=F_0^y=F_0$.
The amplitude and frequency are set to $F_0=2$ and $\mathit{\Omega}=1$, respectively.
The magnitude of the localized spin is taken to be $\vert \bm{S}_i \vert = S = 1$.
We chose numerical values of the Gilbert damping constant $\alpha = 1$ and the time step $\delta t = 0.005$.
We introduce initial fluctuations to the localized spins; the polar angles are uniformly distributed in $[0,\delta\theta]$ with $\delta\theta=0.1$ rad in the initial state~\cite{Koshibae2009,Koshibae2011,Ono2017}.

The real-time dynamics induced by the cw field is shown in Figs.~\ref{fig:timeevol}(a)--\ref{fig:timeevol}(d).
We present the single-particle energy levels $\varepsilon_\nu$, their occupation numbers $n_\nu = \langle \phi_\nu^\dagger \phi_\nu \rangle$, the spin structure factor defined by $S(\bm{q})=N^{-2} \sum_{ij} e^{i\bm{q}(\bm{r}_i-\bm{r}_j)} \bm{S}_i {\cdot} \bm{S}_j$, the Drude weight $\real \sigma_{xx}(0)$, and the optical conductivity $\real \sigma_{xx}(\omega)$, as functions of time $t$.
Figure~\ref{fig:timeevol}(e) shows the optical conductivity averaged during $t=900$ and $1000$.

In the initial state before light irradiation $(t<0)$,
the FM metallic state is realized due to the strong Hund coupling.
The lower (major-spin) and upper (minor-spin) bands are centered at $\pm J$, and the lower band is filled up to $\varepsilon_\nu=-J=-4$.
In an early stage after turning on the cw field, $0<t\lesssim 30$, the localized spin structure and the electron band structure remain unchanged.
The electron momentum distribution is shifted by $\bm{A}(t)$ in the momentum space, which results in coherent ocsillations in $n_\nu$ and $\real \sigma_{xx}(\omega=0)$ with a period of $2\pi/(2\mathit{\Omega})=3.14$ (see Figs.~\ref{fig:timeevol}(a) and \ref{fig:timeevol}(c)).
Then, the FM order characterized by $S(0,0)$ is gradually weaken and the Drude weight diminishes in $30 \lesssim t \lesssim 100$~\footnote{The detail of transient spin structure will be discussed elsewhere.}.
Subsequently, the AFM order characterized by $S(\pi,\pi)$ develops until $t\approx 400$.
Finally, the system reaches the AFM steady state, where the electrons almost uniformly fill the lower band as shown in Fig.~\ref{fig:timeevol}(a).
In the optical conductivity shown in Figs.~\ref{fig:timeevol}(d) and \ref{fig:timeevol}(e), the interband transition peak and its Floquet sidepeaks appear at $\omega = 2J$ and $\omega = 2J \pm n\mathit{\Omega}\ (n=1,2)$, respectively.

The emergence of the AFM steady state is understood in terms of the energetics of the FM and AFM states;
when we assume the ideal FM or AFM spin configuration and the uniform electron distribution in the lower band, the AFM state has the lower energy than the FM state in a wide range of $J$ and $n_e$.
Details are presented in Ref.~\cite{Ono2017}.
A microscopic mechanism of the FM-to-AFM transition and the origin of the polarization dependence (see Figs.~2(d)--2(f) in Ref.~\cite{Ono2017}) are addressed in the following section on the basis of the Floquet Green function method.

\subsection{Magnon spectra in photoirradiated FM metal} \label{sec:spectra}
In this section, we study the magnetic and electronic excitations in the photoirradiated FM metallic state.
We show the time-averaged spectral functions of the magnons and electrons, which are obtained from the Floquet Green function method as follows.
First, we define the inverse of the bare electron Green functions $(\mathcal{G}_{\bm{k}s}^{X,-1})_{mn}(\omega)$ in Eqs.~\eqref{eq:green_retarded_inv} and \eqref{eq:green_keldysh_inv}, and compute $(\mathcal{G}_{\bm{k}s}^{X})_{mn}(\omega)$.
Then, we solve the Dyson equation in Eq.~\eqref{eq:dyson_magnon} for the magnon Green function $(D_{\bm{q}}^{\mathrm{R}})_{mn}(\omega)$ with the retarded selfenergy $(\mathit{\Sigma}_{\bm{q}}^{\mathrm{R}})_{mn}(\omega)$.
The dimension of the Floquet space is set to $N_p=8$--$16$, for which we have numerically confirmed the convergence.
The positive constant $\mathit{\Gamma}$ and the inverse temperature $\beta$ introduced in Eqs.~\eqref{eq:green_retarded_inv}, \eqref{eq:green_keldysh_inv}, and \eqref{eq:dyson_magnon_bare} are set to $\mathit{\Gamma}=0.05$ and $\beta\rightarrow \infty$, respectively.
The chemical potential is chosen to be $\mu=-J$, which keeps the system quarter-filled, i.e., $n_e=0.5$.
The number of sites is taken as $N=32\times 32$ in most of the numerical calculations.

%========================================
\begin{figure}[t]
\includegraphics[width=\columnwidth]{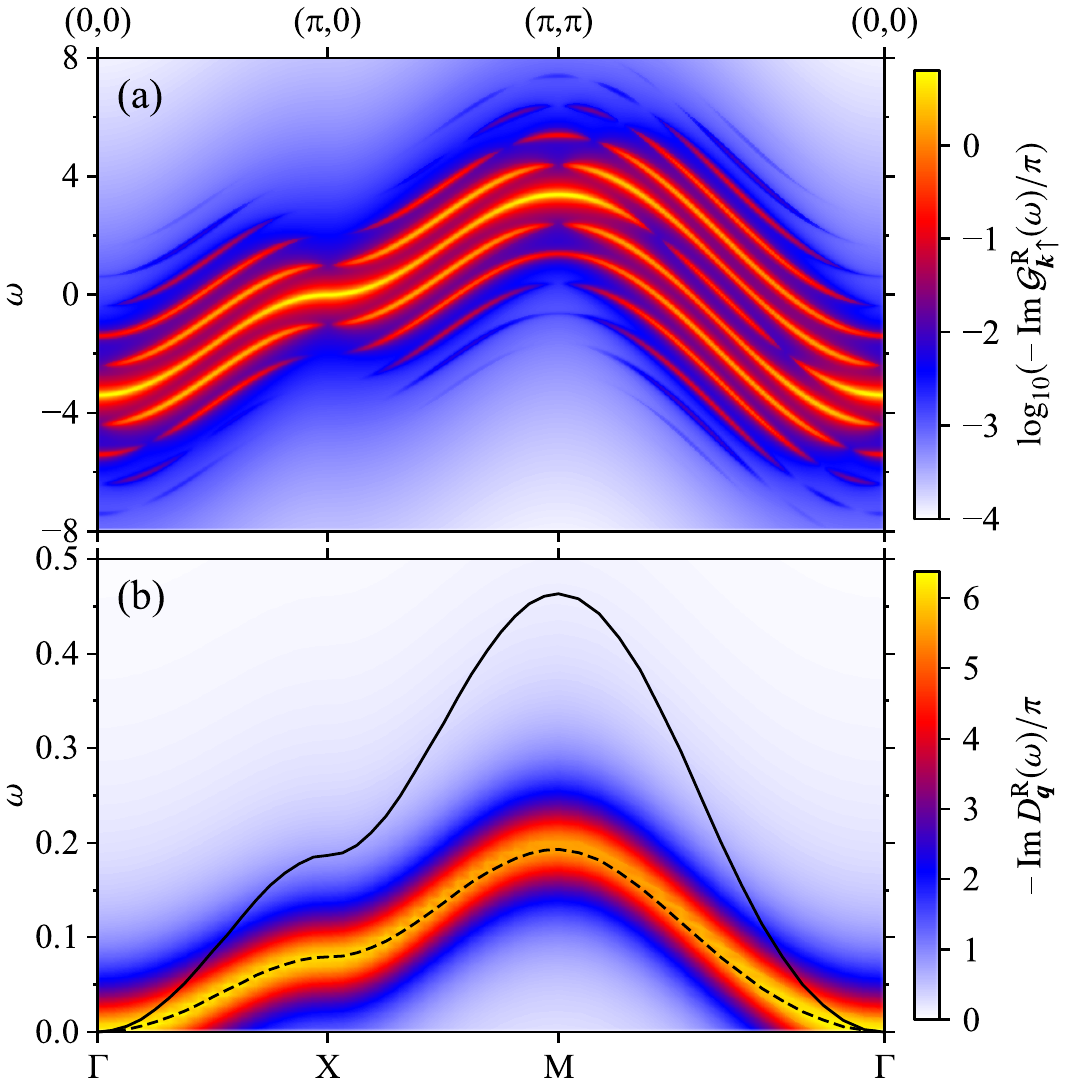}
\caption{The time-averaged spectral functions of (a) the spin-up electrons and (b) the magnons in the cw field with $F_0=0.8$ and $\mathit{\Omega}=1$.
Solid and dashed curves in (b) show the low-energy magnon dispersions obtained from Eq.~\eqref{eq:magnon_dispersion} in the absence and presence of the field, respectively.
The Hund coupling is set to $J=5$.
The number of sites is $N=256\times 256$ in (a) and $N=32\times 32$ in (b).}
\label{fig:spectrum}
\end{figure}
%========================================

Figure~\ref{fig:spectrum} shows the imaginary parts of the time-averaged Green functions, $\mathcal{G}_{\bm{k}\uparrow}^{\mathrm{R}}(\omega)\equiv (\mathcal{G}_{\bm{k}\uparrow}^{\mathrm{R}})_{00}(\omega)$ and $D_{\bm{q}}^{\mathrm{R}}(\omega)\equiv (D_{\bm{q}}^{\mathrm{R}})_{00}(\omega)$, in the cw field with $F_0=0.8$ and $\mathit{\Omega}=1$.
In Fig.~\ref{fig:spectrum}(a) for the electronic band, it is shown that the bandwidth is reduced by a factor of $\mathcal{J}_0(F_0/\mathit{\Omega})\approx 0.85$ and several Floquet sidebands spaced by the frequency $\mathit{\Omega}$ appear.
Modulation of the spectral intensity results from the hybridization between these Floquet bands.
Thus, the magnon spectral function shown in Fig.~\ref{fig:spectrum}(b) shows softening in the whole range of $\bm{q}$.
As comparison, we show the dispersion relation at $F_0=0$ by a solid curve.

%========================================
\begin{figure}[t]
\includegraphics[width=\columnwidth]{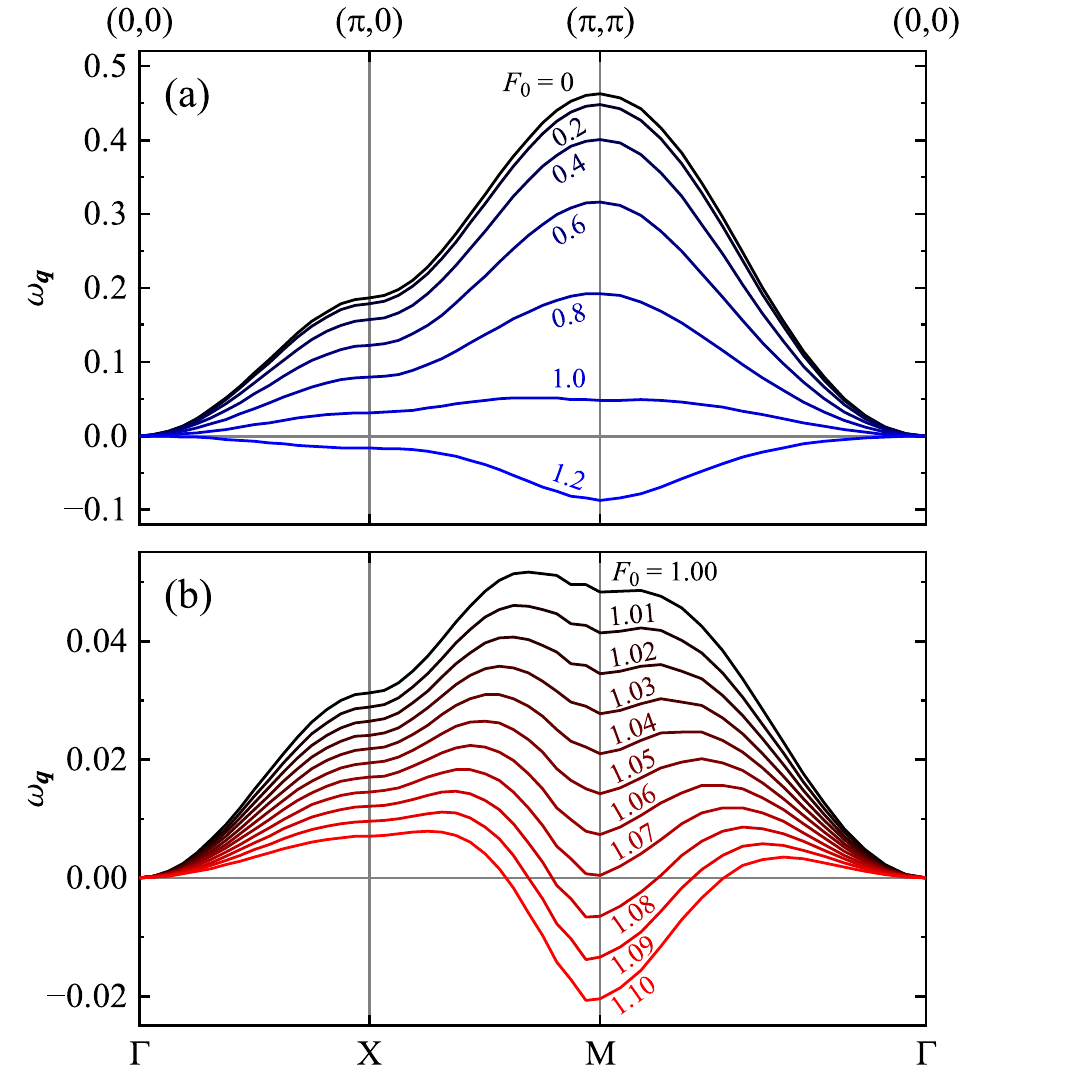}
\caption{The magnon dispersion for (a) $F_0=0.0$--$1.2$, and (b) $F_0=1.0$--$1.1$.
The other parameters are set to $\mathit{\Omega}=1$ and $J=5$.}
\label{fig:dispersion}
\end{figure}
%========================================

We focus on the low-energy magnon dispersion $\omega_{\bm{q}}$, which is approximately given by the equation,
\begin{align}
\real (D_{\bm{q}}^{\mathrm{R},-1})_{00}(\omega_{\bm{q}})=\omega_{\bm{q}} - \omega_{\bm{q}}^{(0)} - \real (\mathit{\Sigma}_{\bm{q}}^{\mathrm{R}})_{00}(\omega_{\bm{q}}) = 0.
\label{eq:magnon_dispersion}
\end{align}
As shown in Fig.~\ref{fig:dispersion}, the magnon dispersion is softened with increasing the electric-field amplitude $F_0$.
In the case of weak irradiation, i.e., $F_0<1.0$ shown in Fig.~\ref{fig:dispersion}(a), the dispersion is similar to that of $F_0=0$ and the bandwidth is reduced.
However, when $F_0\approx 1.0$--$1.1$ shown in Fig.~\ref{fig:dispersion}(b), a dip appears at $(\pi,\pi)$ and the magnon energy at $\bm{q}=(\pi,\pi)$ reaches zero at $F_0\approx 1.07$, which gives rise to instability of the FM state against the AFM one.
This observation is consistent with the results in Ref.~\cite{Ono2017} and Sec.~\ref{sec:real-time}, where the FM-to-AFM transition was demonstrated by the numerical simulation of the real-time dynamics.

%========================================
\begin{figure}[t]
\includegraphics[width=\columnwidth]{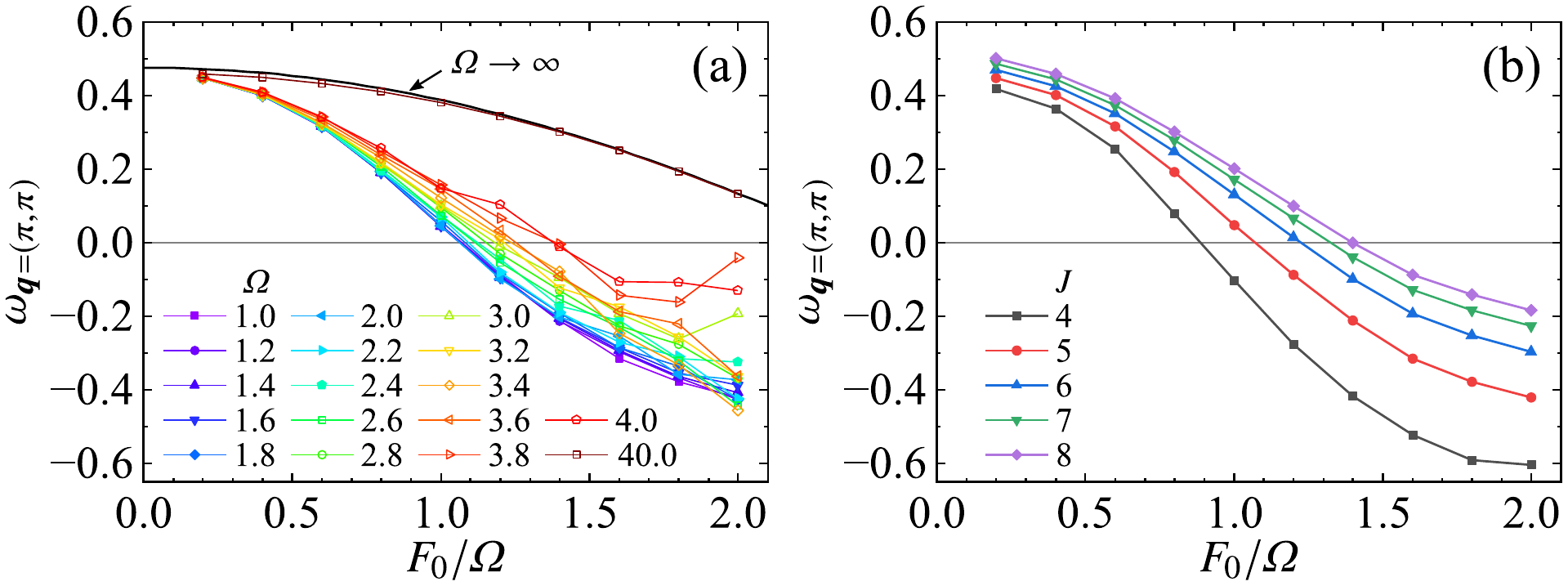}
\caption{The magnon energy at $\bm{q}=(\pi,\pi)$ as a function of the electric-field amplitude divided by the frequency, for different values of (a) the frequency at $J=5$ and (b) the Hund coupling at $\mathit{\Omega}=1$.
A solid curve in (a) shows the magnon dispersion in the limit of $\mathit{\Omega}\rightarrow \infty$.
The chemical potential is set to $\mu=-J$.}
\label{fig:wj-dep}
\end{figure}
%========================================

In Fig.~\ref{fig:wj-dep}, we show the cw frequency $\mathit{\Omega}$ and the Hund coupling $J$ dependences of the magnon energy at the M point, $\omega_{\bm{q} = (\pi,\pi)}$, as functions of $F_0/\mathit{\Omega}$.
In the high-frequency limit ($\mathit{\Omega} \rightarrow \infty$), the off-diagonal components in the Floquet space can be neglected, which simplifies the magnon selfenergy in Eq.~\eqref{eq:selfenergy_retarded} to the following form:
\begin{align}
&(\mathit{\Sigma}_{\bm{q}}^{\mathrm{R}})_{00}(\omega) \notag \\
&= \frac{J}{SN} \sum_{\bm{k}} [f(\varepsilon_{\bm{k}\uparrow})-f(\varepsilon_{\bm{k}+\bm{q}\downarrow})] \notag \\
&\quad\times {\left[ 1 + \frac{2J}{\omega-\mathcal{J}_0(F_0/\mathit{\Omega})(\varepsilon_{\bm{k}+\bm{q}}-\varepsilon_{\bm{k}})-2J+2i\mathit{\Gamma}} \right]}
\label{eq:selfenergy_retarded_dl}
\end{align}
for the linearly-polarized light with $\bm{F}_0=(F_0,F_0)$.
This expression is similar to the equilibrium selfenergy in Eq.~\eqref{eq:selfenergy_retarded_equil} except that the bandwidth is reduced by a factor of $\mathcal{J}_0(F_0/\mathit{\Omega})$ due to the dynamical localization (DL)~\cite{Dunlap1986,Holthaus1992,Grossmann1991,Kayanuma2008,Eckardt2005,Lignier2007}.
The magnon energy at $\bm{q}=(\pi,\pi)$ calculated from the selfenergy in Eq.~\eqref{eq:selfenergy_retarded_dl} is shown as a solid curve in Fig.~\ref{fig:wj-dep}(a), which fits the data for $\mathit{\Omega}=40\ (\gg J,h)$ quite well.
The selfenergy is reduced to $\real(\mathit{\Sigma}_{\bm{q}}^{\rm R})_{00}(\omega=0)\approx 0$ at the zero points of the Bessel function, which means a flat dispersion: $\omega_{\bm{q}}=0$.
Therefore, the dip structure as in Fig.~\ref{fig:dispersion}(b) is not understood only in terms of DL.
On the other hand, in the case where $\mathit{\Omega}$ is comparable to or smaller than $J$ and $h$, the results for $\mathit{\Omega} \leq 4$ in Fig.~\ref{fig:wj-dep}(a) show deviations from the high-frequency curve.
The origin of the deviations is ascribed not only to DL but also to nonequilibrium electron distributions.
It is also found that the magnon energy is scaled to a single curve that crosses the zero energy at $F_0/\mathit{\Omega} \approx 1.1$ for $\mathit{\Omega} \lesssim 2$.
This result is consistent with the fact that the characteristic timescale when the FM order is broken is scaled by $F_0/\mathit{\Omega}$ with finite threshold intensity (see Fig.~3(e) in Ref.~\cite{Ono2017}).
Figure~\ref{fig:wj-dep}(b) shows $\omega_{\bm{q}=(\pi,\pi)}$ for different values of the Hund coupling, indicating that the larger Hund coupling makes the magnon energy higher and thus requires the larger $F_0$ to induce the instability.

%========================================
\begin{figure}[t]
\includegraphics[width=\columnwidth]{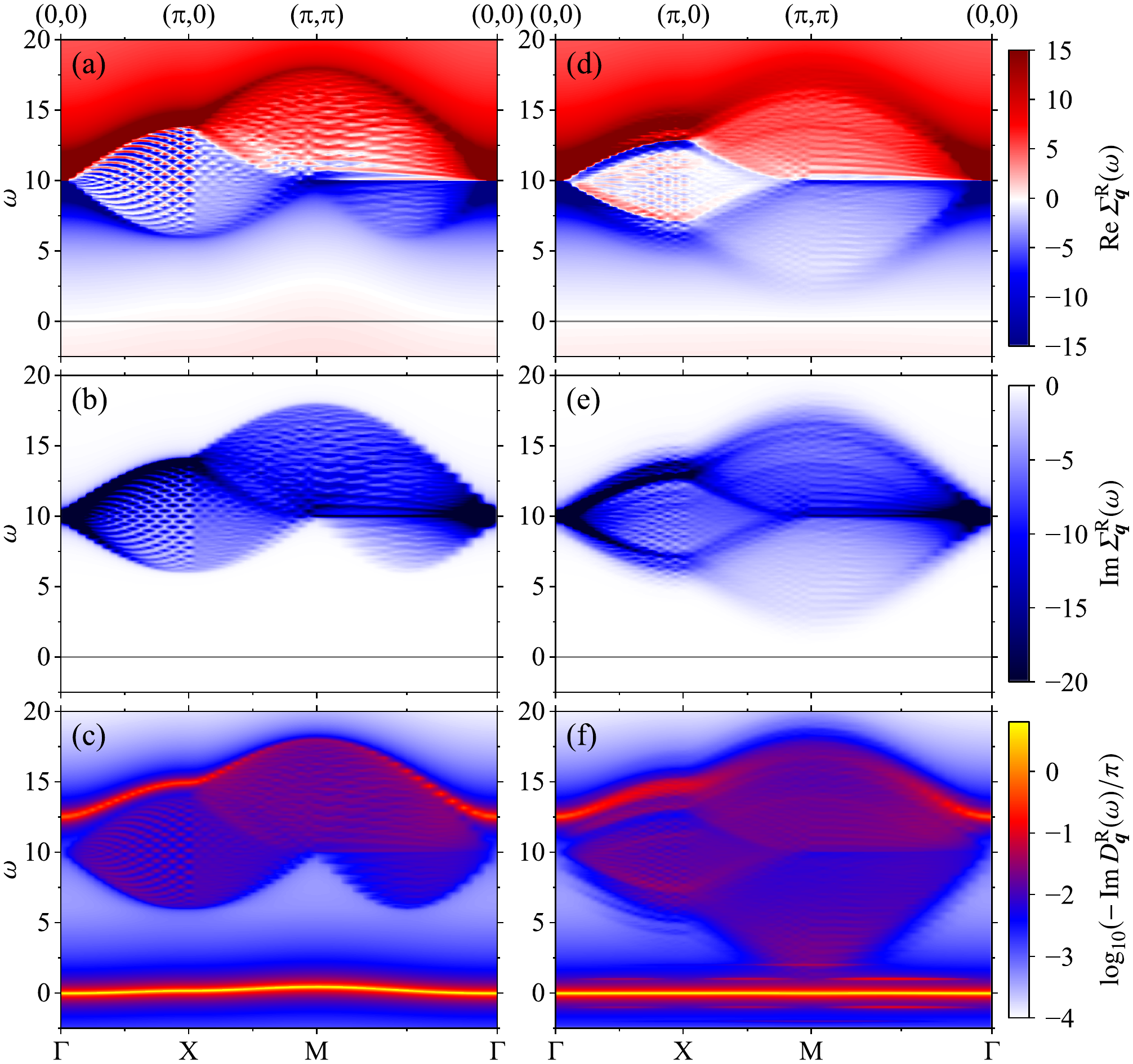}
\caption{Top: the real part of the selfenergy. Middle: the imaginary part of the selfenergy. Bottom: the magnon spectral function.
(a)--(c) are for the equilibrium state ($F_0=0$), and (d)--(f) are for the steady state in the cw field ($F_0=1.07$).
The other parameters are set to $\mathit{\Omega}=1$ and $J=5$.}
\label{fig:spectrum_high}
\end{figure}
%========================================

The magnon selfenergies and spectra in the equilibrium and steady states are shown in Fig.~\ref{fig:spectrum_high}, where the high-energy continuum is seen around $\omega = 2J = 10$, in addition to the low-energy spin-wave excitations.
These high-energy Stoner excitations originate from creations of an electron in the upper band and a hole in the lower band, as diagrammatically represented by Fig.~\ref{fig:diagram}(b).
It is found in Fig.~\ref{fig:spectrum_high}(f) that the high-energy continuum expands to the low-energy region especially at $\bm{q}=(\pi,\pi)$ with increasing $F_0$.
The energy range of the continuum excitations reflects the electron distribution and the electron joint density of states,
since the imaginary part of the selfenergy in the equilibrium state obtained from Eq.~\eqref{eq:selfenergy_retarded_equil} is given by
\begin{align}
\imaginary \mathit{\Sigma}_{\bm{q}}^{\mathrm{R}}(\omega)
\approx -\frac{2J^2}{SN} \sum_{\bm{k}} f(\varepsilon_{\bm{k}\uparrow}) \pi \delta(\omega-(\varepsilon_{\bm{k}+\bm{q}\downarrow}-\varepsilon_{\bm{k}\uparrow})),
\label{eq:selfenergy_retarded_equil_im}
\end{align}
where $f(\varepsilon_{\bm{k}+\bm{q}\downarrow}) \approx 0$ is taken.

%========================================
\begin{figure}[t]
\includegraphics[width=0.9\columnwidth]{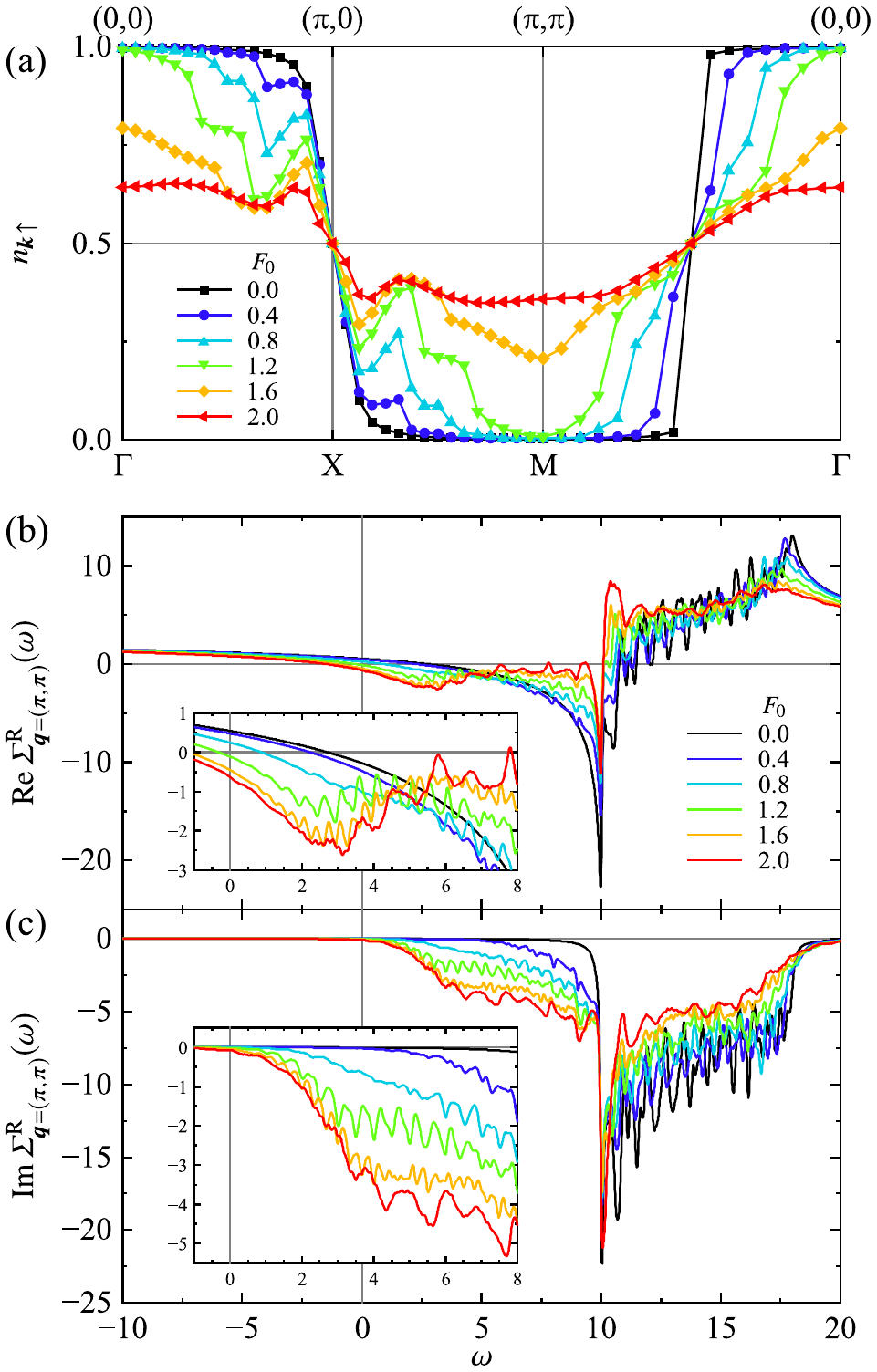}
\caption{(a) The momentum distribution of the spin-up electrons.
(b) The real part and (c) the imaginary part of the time-averaged selfenergy at $\bm{q}=(\pi,\pi)$.
Insets of (b) and (c) are the extensions for the low-energy region.
We chose $\mathit{\Omega}=1$ and $J=5$.
}
\label{fig:selfenergy}
\end{figure}
%========================================

The mechanism of this magnon softening is ascribed to the electron distribution in the light-induced steady state and is understood on the basis of the equilibrium magnon selfenergy in Eq.~\eqref{eq:selfenergy_retarded_equil}.
The momentum distribution function of the spin-up electrons defined by
\begin{align}
n_{\bm{k}\uparrow} = \int_{-\infty}^{\infty} \frac{d\omega}{2\pi} \imaginary G_{\bm{k}\uparrow}^{<}(\omega)
\end{align}
is shown in Fig.~\ref{fig:selfenergy}(a), where the distribution is changed into a uniform distribution, $n_{\bm{k}\uparrow}=n_e$, with increasing $F_0$.
Assuming that the expression of the equilibrium selfenergy in Eq.~\eqref{eq:selfenergy_retarded_equil} is valid in the steady states, we notice that the transverse component of the selfenergy $\mathit{\Sigma}_{2}$ reduces the magnon energy as
\begin{align}
\real \mathit{\Sigma}_{2,\bm{q}}^{\mathrm{R}}(\omega=0) \approx -\frac{2J^2}{SN} \sum_{\bm{k}} \frac{ n_{\bm{k}\uparrow} }{\varepsilon_{\bm{k}+\bm{q}}-\varepsilon_{\bm{k}}+2J} < 0.
\label{eq:selfenergy2_retarded_equil2}
\end{align}
Here, we replace the Fermi--Dirac function $f(\varepsilon_{\bm{k}s})$ by $n_{\bm{k}s}$, neglect $n_{\bm{k}+\bm{q}\downarrow}$, and assume that $2J$ is larger than the bandwidth ($=8$ in the present model) \footnote{The condition $2J>8$ is not always necessary for $\real \mathit{\Sigma}_{2,\bm{q}}^{\mathrm{R}}(0) < 0$.
However, when $2J$ is smaller than the bandwidth, the imaginary part of the selfenergy in Eq.~\eqref{eq:selfenergy_retarded_equil_im} is finite at $\omega=0$ and makes the low-energy magnons ill-defined.}.
In the perturbative process in $\mathit{\Sigma}_{2}$ representing the Stoner excitation, the change in the electron distribution increases the available momentum phase space governed by $n_{\bm{k}\uparrow}$ in the summation over $\bm{k}$, in Eq.~\eqref{eq:selfenergy2_retarded_equil2}.
This means that, under the photoirradiation, the low-energy electron-hole excitations are allowed because of the change in $n_{\bm{k}\uparrow}$ and generate a new scattering continuum in $\omega < 2J$.
The energy gain by the Stoner processes takes its maximum at $\bm{q}=(\pi,\pi)$ in the present square lattice, since the energy denominator $\varepsilon_{\bm{k}+\bm{q}}-\varepsilon_{\bm{k}}+2J=2J-2\varepsilon_{\bm{k}}>0$ takes its minimum.
Figures~\ref{fig:selfenergy}(b) and \ref{fig:selfenergy}(c) show the time-averaged selfenergy at $\bm{q}=(\pi,\pi)$ for different values of the cw amplitude.
In the low-energy region around $\omega\approx 0$, the real part decreases with increasing $F_0$ and the imaginary part is almost zero for $F_0 \lesssim 1.2$.
These results mean that the magnon at $\bm{q}=(\pi,\pi)$ remains well-defined and is softened by the photoirradiation.

%========================================
\begin{figure}[t]
\includegraphics[width=0.9\columnwidth]{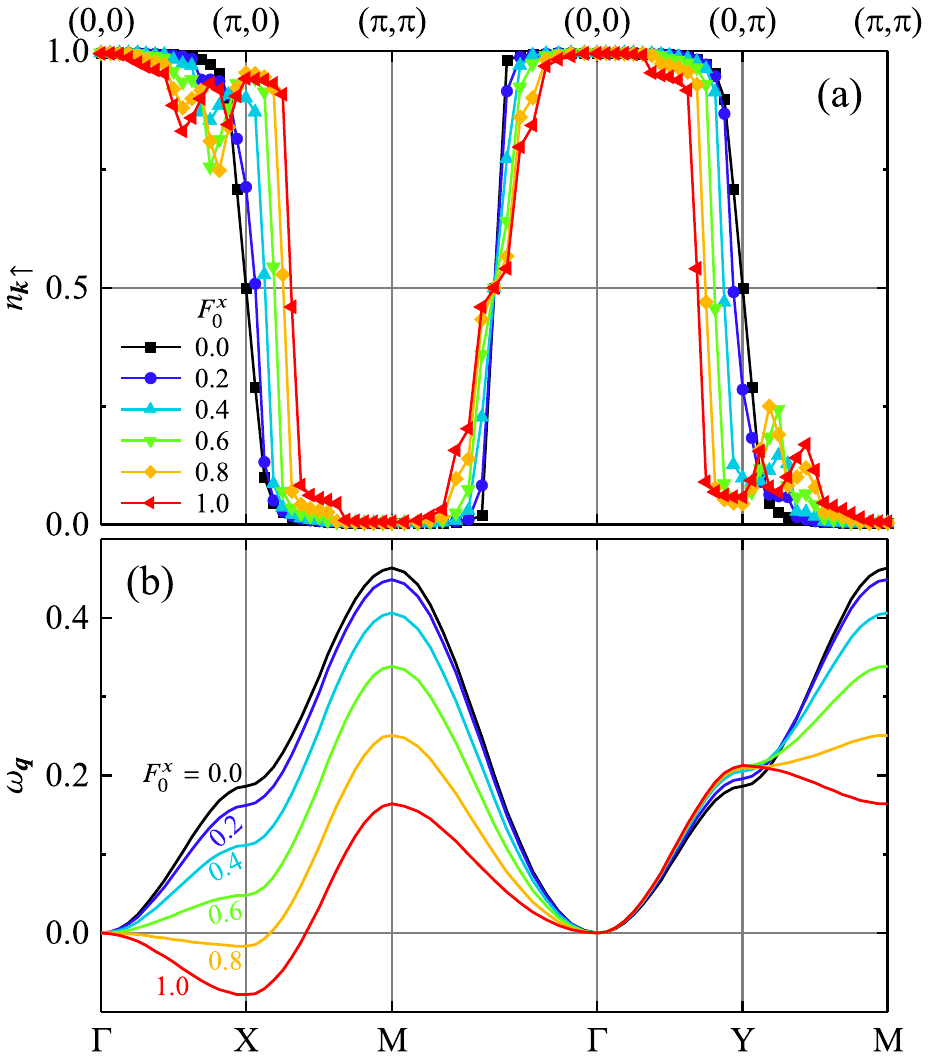}
\caption{(a) The momentum distribution $n_{\bm{k}\uparrow}$, and (b) the magnon dispersion $\omega_{\bm{q}}$ with $\bm{F}_0=(\sqrt{2}F_0^x,0)$.
Other parameter values are $\mathit{\Omega}=1$ and $J=5$.}
\label{fig:dispersion_pol}
\end{figure}
%========================================

Finally, we discuss the polarization dependence of the softening.
The momentum distribution function and the magnon dispersion are shown in Fig.~\ref{fig:dispersion_pol} for different amplitudes.
We set the linearly polarized light along the $x$ direction as $\bm{F}_0=(\sqrt{2}F_0^x,0)$.
Equations~\eqref{eq:floquetband_even} and \eqref{eq:floquetband_odd} are changed to
%\begin{subequations}
\begin{align}
\bar{\varepsilon}_{mn,\bm{k}} = -2\mathcal{J}_{m-n}(\sqrt{2}F_0^x/\mathit{\Omega}) \cos k_x - 2 \cos k_y
\end{align}
for $m-n = 0 \bmod 2$, and
\begin{align}
\bar{\varepsilon}_{mn,\bm{k}} = -2i\mathcal{J}_{m-n}(\sqrt{2}F_0^x/\mathit{\Omega}) \sin k_x - 2 i \sin k_y
\end{align}%\end{subequations}
for $m-n = 1 \bmod 2$, respectively.
It is shown that the momentum distribution decreases along the $\Gamma$--Y line and increases along the X--M line with increasing $F_0^x$.
Thus, the magnon momentum that minimizes the energy denominator in Eq.~\eqref{eq:selfenergy2_retarded_equil2} under the condition of $n_{\bm{k}\uparrow}>0$ is given by $\bm{q}\approx (\pi,0)$.
Consequently, the magnon at $\bm{q}=(\pi,0)$ is softened rather than that at $\bm{q}=(\pi,\pi)$.
This is consistent with the polarization dependence of the transient spin structure shown in Fig.~2(d)--2(f) in Ref.~\cite{Ono2017}.
As for the circularly-polarized light, no major differences from the case of $\bm{F}_0=(F_0,F_0)$ are observed except for the dip structure seen in Fig.~\ref{fig:dispersion}(b) (not shown).
This is because the electric field does not couple directly to the electron spins in the present model, where the spin-orbit coupling is not taken into account.

\section{Summary}
\label{sec:summary}
We have studied the photoinduced dynamics in the itinerant magnet described by the DE model.
It is found that the initial FM metallic state is changed to the AFM state by the cw field, which is in sharp contrast to the well-known AFM-to-FM transition due to the photocarrier injection.
We presented formulation for the transient optical conductivity spectra by extending the formalism based on nonequilibrium Green function~\cite{Eckstein2008} to an inhomogeneous system.
It is found that, in the photoinduced AFM steady state, the interband excitation peak at $\omega = 2J$ and the Floquet sidepeaks at $\omega = 2J\pm n\mathit{\Omega}\ (n=1,2,\dots)$ appear.
These are available to identify the FM-to-AFM transition proposed in the present paper.

We also investigated the magnetic excitation properties in the FM metal in the cw light by using the Floquet Green function method.
The magnon Green function is calculated in the perturbative expansion with respect to the Hund coupling, where the Hartree-type and bubble-type diagrams are taken into account.
It is found that, with increasing the cw amplitude, the magnon dispersion is softened in the whole momentum range, and the dip structure appears at $\bm{q}=(\pi,\pi)$ in the square lattice for $F_0\geq 1.0$.
This implies that the FM state is unstable due to the photoirradiation and is transformed into the AFM state at the finite cw amplitude.
In the low-frequency regime $\mathit{\Omega}\lesssim 2$, the magnon energy at $\bm{q}=(\pi,\pi)$ is scaled to the single curve and is lower than that in the high-frequency limit $\mathit{\Omega}\rightarrow \infty$.
These observations based on the Floquet Green function method are consistent with the results by the real-time simulation in Ref.~\cite{Ono2017}, and reveal the microscopic mechanism of the FM-to-AFM transition as follows.
In the FM steady state, the electron momentum distribution is modulated by the cw field, which enhances the low-energy Stoner excitation and reduces the magnon energy.
The nonequilibrium electron distribution induced by the cw field plays a crucial role on the softening of the magnons and the appearance of the dip structure in the magnon dispersion.
This is beyond the DL effect that appears in the high-frequency limit and leads to the monotonic reduction of the magnon bandwidth.

\begin{acknowledgments}
This work was supported by JSPS KAKENHI Grant No.~15H02100, No.~17H02916, No.~18H05208, and No.~18J10246.
The computation in this work has been done using the facilities of the Supercomputer Center, the Institute for Solid State Physics, the University of Tokyo.
\end{acknowledgments}

\appendix

\section{Keldysh formalism}
\label{sec:keldysh}

\begin{figure}[b]\centering
\includegraphics[scale=1]{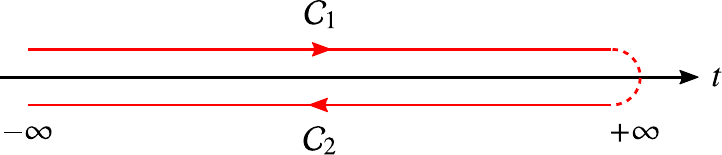}
\caption{The Schwinger--Keldysh contour: $\mathcal{C}=\mathcal{C}_1+\mathcal{C}_2$.}
\label{fig:contour}
\end{figure}

We briefly introduce the Keldysh formalism and the contour-ordered Green function (see, e.g., Refs.~\cite{Rammer1986,Kita2010,Aoki2014,Rammer2007,Altland2010,Kamenev2011,Stefanucci2013,Citro2018} for details).
Let $\vert \Psi(-\infty) \rangle \equiv \vert \Psi_0 \rangle$ be an initial state.
The expectation value of an operator $O({t})$ at time ${t}$ is represented as
\begin{align}
\langle O({t}) \rangle &\equiv \langle \Psi_0 \vert O({t}) \vert \Psi_0 \rangle \notag \\
&= \langle \Psi_0 \vert U^\dagger({t},-\infty) O U({t},-\infty) \vert \Psi_0 \rangle,
\label{eq:expectationvalue}
\end{align}
where $O=O(-\infty)$, and the unitary operator $U$ is given by
\begin{align}
U({t},{t}') = \begin{cases}
\mathcal{T}_{\mathcal{C}_1} \exp \left[ -i \int_{{t}'}^{{t}} d\bar{{t}}\, H(\bar{{t}}) \right] & ({t}>{t}'), \\[2mm]
\mathcal{T}_{\mathcal{C}_2} \exp \left[ -i \int_{{t}'}^{{t}} d\bar{{t}}\, H(\bar{{t}}) \right] & ({t}<{t}').
\end{cases}
\label{eq:unitary}
\end{align}
The symbol $\mathcal{T}_{\mathcal{C}_1}$ ($\mathcal{T}_{\mathcal{C}_2}$) represents a (anti-)time-ordered operator.
Using $U({t},{t}')U({t}',{t}'')=U({t},{t}'')$ and $U^\dagger({t},{t}')=U({t}',{t})$, the expectation value in Eq.~\eqref{eq:expectationvalue} is written as
\begin{align}
\langle O({t}) \rangle
= \langle \Psi_0 \vert \mathcal{T}_{\mathcal{C}} \exp \left(-i\int_{\mathcal{C}} d\bar{{t}}\, H(\bar{{t}})\right) O({t}) \vert \Psi_0 \rangle,
\label{eq:expectationvalue_contour}
\end{align}
where $\mathcal{T}_{\mathcal{C}}$ is the contour-ordered operator defined on the Schwinger--Keldysh contour $\mathcal{C}$ depicted in Fig.~\ref{fig:contour}.

%========================================
%========================================

When the Hamiltnian $H$ is divided into the non-interacting part $H_0=\sum_\nu \varepsilon_\nu \psi_\nu^\dagger \psi_\nu$ and the perturbative part $V$ as $H=H_0+V$, it is useful to introduce the interaction picture to perform a perturbative expansion.
By introducing a time-evolution operator $U_0({t},{t}')$ as the non-interacting counterpart of Eq.~\eqref{eq:unitary}, in which $H$ is replaced with $H_0$, we define the $S$-matrix as
\begin{align}
S_{\mathcal{C}} = \mathcal{T}_{\mathcal{C}} \exp {\left[ -i \int_\mathcal{C} d\bar{{t}}\, V_\mathrm{I}(\bar{{t}}) \right]},
\label{eq:smatrix}
\end{align}
where $V_{\mathrm{I}}({t}) = U_0^\dagger({t},-\infty)V({t})U_0({t},-\infty)$ is the perturbation in the interacting picture.
The expectation value of $O({t})$ in Eq.~\eqref{eq:expectationvalue_contour} is given by
\begin{align}
\langle O({t}) \rangle = \langle \Psi_0 \vert \mathcal{T}_{\mathcal{C}} S_{\mathcal{C}} O_{\mathrm{I}}({t}) \vert \Psi_0 \rangle,
\end{align}
with $O_{\mathrm{I}}({t}) = U_0^\dagger({t},-\infty) O U_0({t},-\infty)$.

We introduce the contour-ordered Green function $G(t,t')$ as
\begin{align}
iG_{\mu\nu}({t},{t}') &= \langle \mathcal{T}_{\mathcal{C}} \psi_\mu({t}) \psi_\nu^\dagger({t}') \rangle
= \langle \mathcal{T}_{\mathcal{C}} S_{\mathcal{C}} \psi_{\mathrm{I}\mu}({t}) \psi_{\mathrm{I}\nu}^\dagger({t}') \rangle,
\label{eq:contourgreenfunc}
\end{align}
where $\psi_\nu^\dagger$ is a creation operator of a boson or fermion with a quantum number $\nu$.
The contour-ordered Green function is expressed in a matrix form as
\begin{align}
\hat{G}_{\mu\nu}(t,t') = {\begin{bmatrix} G^{11}_{\mu\nu}(t,t') & G^{12}_{\mu\nu}(t,t') \\ G^{21}_{\mu\nu}(t,t') & G^{22}_{\mu\nu}(t,t') \end{bmatrix}},
\end{align}
where the superscripts, $1$ and $2$, denote the branch of the contour $\mathcal{C}$ to which the time variables belong.
Since the contour-ordered function $G$ satisfies the equation:
$
G^{11}+G^{22}=G^{12}+G^{21},
$
the redundancy is eliminated by the Keldysh rotation given by
\begin{align}
\hat{G} \mapsto \tilde{G} = L\sigma^z \hat{G} L^\dagger
= {\begin{bmatrix} G^{\mathrm{R}} & G^{\mathrm{K}} \\ 0 & G^{\mathrm{A}} \end{bmatrix}},
\end{align}
where
\begin{align}
L=\frac{1}{\sqrt{2}} {\begin{bmatrix} 1 & -1 \\ 1 & 1 \end{bmatrix}}
\end{align}
is a unitary matrix and $G^{\mathrm{R}}$, $G^{\mathrm{A}}$, and $G^{\mathrm{K}}$ are the retarded, advanced, and Keldysh Green functions, respectively. The lesser Green function $G^{<}=G^{12}$ and the greater one $G^{>}=G^{21}$ are given by $G^{<} = (G^\mathrm{K}-G^\mathrm{R}+G^\mathrm{A})/2$ and $G^{>} = (G^\mathrm{K}+G^\mathrm{R}-G^\mathrm{A})/2$.

The equation of motion (Dyson equation) of the contour-ordered Green function $\hat{G}$ is given by
\begin{align}
[\hat{\mathcal{G}}^{-1} \circ \hat{G}]({t},{t}')
= \sigma^z \delta({t}-{t}') + [\hat{\mathit{\Sigma}} \circ \hat{G}]({t},{t}'),
\label{eq:eom1}
\end{align}
where $\hat{\mathcal{G}}_{\mu\nu}^{-1}({t},{t}') = \delta_{\mu\nu} \sigma^z \delta({t}-{t}')(i\partial_{t} - \varepsilon_\nu)$ is the inverse of the bare Green function and $\hat{\mathit{\Sigma}}$ is the selfenergy.
Here, we omit a summation over the quantum numbers.
The symbol `$\circ$' in Eq.~\eqref{eq:eom1} represents the convolution defined by
\begin{align}
[\hat{A} \circ \hat{B}]({t},{t}') &= \int_{-\infty}^{\infty} d\bar{{t}}\, \hat{A}({t},\bar{{t}}) \sigma^z \hat{B}(\bar{{t}},{t}') ,
\end{align}
for two-time functions $\hat{A}$ and $\hat{B}$.
The Keldysh rotation transforms Eq.~\eqref{eq:eom1} into
\begin{align}
[(\tilde{\mathcal{G}}^{-1}-\tilde{\mathit{\Sigma}}) \circ \tilde{G}]({t},{t}') = \delta({t}-{t}') ,
\label{eq:eom2}
\end{align}
with
\begin{align}
[\tilde{A} \circ \tilde{B}]({t},{t}')
= \int_{-\infty}^{\infty} d\bar{{t}}\, \tilde{A}({t},\bar{{t}}) \tilde{B}(\bar{{t}},{t}').
\end{align}
We identify $\tilde{\mathcal{G}}^{-1}-\tilde{\mathit{\Sigma}}$ in Eq.~\eqref{eq:eom2} as inverse of the full Green function $\tilde{G}^{-1}$.
Finally, the Dyson equation is written as
\begin{align}
\begin{bmatrix} G^\mathrm{R} & G^\mathrm{K} \\ 0 & G^\mathrm{A} \end{bmatrix}^{-1}
= \begin{bmatrix} \mathcal{G}^\mathrm{R} & \mathcal{G}^\mathrm{K} \\ 0 & \mathcal{G}^\mathrm{A} \end{bmatrix}^{-1} - {\begin{bmatrix} \mathit{\Sigma}^\mathrm{R} & \mathit{\Sigma}^\mathrm{K} \\ 0 & \mathit{\Sigma}^\mathrm{A} \end{bmatrix}},
\label{eq:dyson}
\end{align}
which is an integro-differential equation with respect to time.

In the case of the non-interacting fermionic Hamiltonian given by $H_0=\sum_\nu \varepsilon_\nu \psi_\nu^\dagger \psi_\nu$, the bare Green functions are written as
%\begin{subequations}\label{eq:free_twotime}
\begin{align}
\mathcal{G}_\nu^{\mathrm{R}}(t,t') &= -i\theta(t-t') e^{-i\varepsilon_\nu (t-t')}
= \mathcal{G}_\nu^{\mathrm{A}}(t',t)^*, \label{eq:free_twotime_retarded} \\
\mathcal{G}_\nu^{\mathrm{K}}(t,t') &= - i (1-2n_\nu^{(0)}) e^{-i\varepsilon_\nu (t-t')}, \label{eq:free_twotime_keldysh} \\
\mathcal{G}_\nu^{<}(t,t') &= i n_\nu^{(0)} e^{-i\varepsilon_\nu (t-t')}, \label{eq:free_twotime_lesser}
\end{align}%\end{subequations}
where $n_\nu^{(0)}=\langle \psi_\nu^\dagger \psi_\nu \rangle$ is the initial distribution function.

\section{Floquet Green function}
\label{sec:floquet-green}
The Floquet Green function method~\cite{Aoki2014,Tsuji2008,Oka2009,Tsuji2009,Mikami2016,Murakami2017,Morimoto2016,Lee2017,Qin2018} efficiently describes the nonequilibrium steady states driven by a time-periodic external field, in which the Green function $G({t},{t}')$ satisfies the relation,
\begin{align}
G^X({t}+T,{t}'+T) = G^X({t},{t}') \quad (X=\mathrm{R},\mathrm{A},\mathrm{K}) ,
\end{align}
with a period of $T=2\pi/\mathit{\Omega}$.
Owing to the periodicity, we introduce the Floquet representation of a two-time function $A(t,t')$, which is called the Floquet Green function in the case of $A(t,t')=G^X(t,t')$, and its inverse transformation as
\begin{align}
A_{mn}(\omega)
&= \int_0^T \frac{dt_a}{T} \int_{-\infty}^{\infty} dt_r\, e^{i(\omega+m\mathit{\Omega}){t}-i(\omega+n\mathit{\Omega}){t}'} A({t},{t}'),
\label{eq:floquet-twotime} \\
A({t},{t}')
&= \sum_{n} \int_{-\infty}^{\infty} \frac{d\omega}{2\pi}\, e^{-in\mathit{\Omega} t_a} e^{-i(\omega+(n/2)\mathit{\Omega})t_r} A_{n,0}(\omega) ,
\label{eq:twotime-floquet}
\end{align}
respectively.
Here, the indices $m$ and $n$ are integer numbers, and $t_a=({t}+{t}')/2$ and $t_r={t}-{t}'$.
Equation~\eqref{eq:floquet-twotime} leads to redundancy of $A_{mn}(\omega)=A_{m+l,n+l}(\omega-l\mathit{\Omega})$.
In the Floquet representation, the Dyson equation given in Eq.~\eqref{eq:dyson} is simplified to a set of the algebraic equations;
the retarded and Keldysh components of $\tilde{G}$ are given by
%\begin{subequations}
\begin{align}
&(G^{\mathrm{R},-1})_{mn}(\omega) = (\mathcal{G}^{\mathrm{R},-1})_{mn}(\omega) - \mathit{\Sigma}^\mathrm{R}_{mn}(\omega), \label{eq:dyson_floquet1} \\
&G^{\mathrm{K}}_{mn}(\omega) = -[G^{\mathrm{R}} (\mathcal{G}^{\mathrm{K},-1}-\mathit{\Sigma}^\mathrm{K}) G^{\mathrm{A}}]_{mn}(\omega), \label{eq:dyson_floquet2}
\end{align}%\end{subequations}
respectively,
where $(G^{X,-1})_{mn}$ is the inverse matrix of $G^X_{mn}$.
Since the inverse of the bare Keldysh Green function, $\mathcal{G}^{\mathrm{K},-1}$, is proportional to an infinitesimal constant, the full Keldysh Green function in Eq.~\eqref{eq:dyson_floquet2} is given by $G_{mn}^{\mathrm{K}}(\omega) = (G^{\mathrm{R}}\mathit{\Sigma}^{\mathrm{K}}G^{\mathrm{A}})_{mn}(\omega)$.
To stabilize nonequilibrium steady states in an external field, we introduce a heat bath with constant density of states, which are incorporated via the following selfenergies:
%\begin{subequations}
\begin{align}
(\mathit{\Sigma}^{\mathrm{R}}_{\text{bath}})_{mn}(\omega) &= -i\delta_{mn}\mathit{\Gamma}, \\
(\mathit{\Sigma}^{\mathrm{K}}_{\text{bath}})_{mn}(\omega) &= -2i\delta_{mn}(1-2f(\omega+n\mathit{\Omega}))\mathit{\Gamma},
\end{align}%\end{subequations}
where $\mathit{\Gamma} \ (>0)$ is coupling strength between the system and the bath, and $f(\omega)$ is the Fermi--Dirac distribution function~\cite{Aoki2014,Tsuji2009,Mikami2016,Murakami2017}.
In this paper, we merge these bath selfenergies into the bare Green functions in Eqs.~\eqref{eq:green_retarded_inv} and \eqref{eq:green_keldysh_inv}.

We define the Wigner representation as
\begin{align}
A(\omega,t_a) &= \int_{-\infty}^{\infty} dt_r\, e^{i\omega t_r} A({t},{t}') \label{eq:wigner-twotime} \\
&= \sum_{\mathclap{n}} e^{-in\mathit{\Omega} t_a} A_{n,0}{\left(\omega-\frac{n\mathit{\Omega}}{2}\right)}, \label{eq:wigner-floquet}
\end{align}
which is useful to investigate the dynamical properties in the nonequilibrium systems at time $t_a$.
In particular, the time average of $A(\omega,t_a)$ is represented by
\begin{align}
A(\omega) \equiv \int_0^T \frac{dt_a}{T}\, A(\omega,t_a) = A_{nn}(\omega-n\mathit{\Omega}),
\end{align}
where $n$ is chosen such that $\omega-n\mathit{\Omega}\in (-\mathit{\Omega}/2,\mathit{\Omega}/2]$.

\section{Response function} \label{sec:responsefunction}
In this section, we derive general expressions of two-body response functions, following the formalism for the optical conductivity that was presented in Ref.~\cite{Eckstein2008}.
Let us consider a response of a one-body operator defined by
\begin{align}
O^\alpha({t}) = \sum_{\mu\nu} O_{\mu\nu}^\alpha({t}) \psi_\mu^\dagger \psi_\nu
\label{eq:response_operator}
\end{align}
to an external field $f^\alpha({t})$, whose coupling Hamiltonian is given by
\begin{align}
V_{\text{ext}}({t}) = - \sum_{\mu\nu} \mathcal{F}_{\mu\nu}({t}) \psi_\mu^\dagger \psi_\nu.
\label{eq:response_coupling}
\end{align}
Here, $\mu$ and $\nu$ are the indices for quantum numbers, $\alpha$ represents a physical index such as the Cartesian coordinate $\alpha=x,y,z$ and momentum transfer $\alpha=\bm{q}$, and $\mathcal{F}$ is a functional of the external field $f$.
A response function (susceptibility) is defined by a functional derivative
\begin{align}
\chi_{\alpha\beta} = \frac{\delta\langle O^\alpha({t})\rangle}{\delta f^\beta({t}')}.
\label{eq:response_def}
\end{align}
The expectation value $\langle O^\alpha({t}) \rangle$ is written in terms of the lesser Green function $G_{\mu\nu}^{<}({t},{t}')=i\langle\psi_\nu^\dagger({t}')\psi_\mu({t})\rangle$ for fermions as
\begin{align}
\langle O^\alpha({t}) \rangle
= -i \sum_{\mu\nu} O_{\mu\nu}^\alpha({t}) G_{\nu\mu}^{<}({t},{t}).
\label{eq:response_expectationvalue}
\end{align}
The derivative of Eq.~\eqref{eq:response_expectationvalue} with respect to the external field $f^\beta({t}')$ yields
%\begin{subequations}
\begin{align}
\chi_{\alpha\beta}({t},{t}')
&= \chi_{\alpha\beta}^{\mathrm{dia}}({t},{t}') + \chi_{\alpha\beta}^{\mathrm{pm}}({t},{t}'), \\
\chi_{\alpha\beta}^{\mathrm{dia}}({t},{t}')
&= -i \sum_{\mu\nu} \frac{\delta O_{\mu\nu}^\alpha({t})}{\delta f^\beta({t}')} G_{\nu\mu}^{<}({t},{t}), \label{eq:response_dia} \\
\chi_{\alpha\beta}^{\mathrm{pm}}({t},{t}')
&= -i \sum_{\mu\nu} O_{\mu\nu}^\alpha({t}) \frac{\delta G_{\nu\mu}^{<}({t},{t})}{\delta f^\beta({t}')} ,
\label{eq:response_pm}
\end{align}%\end{subequations}
where $\chi^{\mathrm{dia}}$ and $\chi^{\mathrm{pm}}$ describe the ``diamagnetic'' and ``paramagnetic'' responses, respectively.
We consider the full Green function given by
\begin{align}
\hat{G}^{-1}_{\mu\nu}({t},{t}')
&= \hat{\mathcal{G}}^{-1}_{\mu\nu}({t},{t}') + \sigma^z \delta(t-t') \mathcal{F}_{\mu\nu}({t}) - \hat{\mathit{\Sigma}}_{\mu\nu}({t},{t}'),
\end{align}
with $\hat{\mathcal{G}}^{-1}_{\mu\nu}({t},{t}') = \delta_{\mu\nu} \sigma^z \delta(t-t') (i\partial_{t} - \varepsilon_\nu)$ being the bare Green function.
The derivative in Eq.~\eqref{eq:response_pm} is expressed as
\begin{widetext}\begin{align}
\frac{\delta G_{\nu\mu}^{<}({t},{t})}{\delta f^\beta({t}')}
&= -\sum_{\kappa\lambda} \int_{-\infty}^{\infty} d\bar{{t}}\, \biggl[G_{\nu\kappa}^{\mathrm{R}}({t},\bar{{t}}) \frac{\delta \mathcal{F}_{\kappa\lambda}(\bar{{t}})}{\delta f^\beta({t}')} G_{\lambda\mu}^{<}(\bar{{t}},{t})
+ G_{\nu\kappa}^{<}({t},\bar{{t}}) \frac{\delta \mathcal{F}_{\kappa\lambda}(\bar{{t}})}{\delta f^\beta({t}')} G_{\lambda\mu}^{\mathrm{A}}(\bar{{t}},{t}) \biggr],
\label{eq:response_pm_derivative}
\end{align}
where we take the variation of the Dyson equation~\eqref{eq:eom1} with respect to $f$ and neglect the vertex correction which arises from $\delta \mathit{\Sigma}/\delta f^\beta$~\cite{Eckstein2008,Tsuji2009,Aoki2014,Tsuji2015,Murakami2016,Tsuji2016}.
The explicit forms of $O_{\mu\nu}^\alpha$ and $\mathcal{F}_{\mu\nu}$ are required for further calculations.
However, in most cases, the coupling $\mathcal{F}(\bar{t})$ depends only on the external field $f^\beta(t')$ at time $t'=\bar{t}$, i.e., $\delta \mathcal{F}(\bar{t})/\delta f^\beta(t')\propto \delta(\bar{t}-t')$, which leads to
\begin{align}
\chi_{\alpha\beta}^{\mathrm{pm}}(t,t')
&= i \Tr \biggl[ O^{\alpha}(t) G^{\mathrm{R}}(t,t') \frac{\partial \mathcal{F}(t')}{\partial f^\beta(t')} G^{<}(t',t)
+ O^{\alpha}(t) G^{<}(t,t') \frac{\partial \mathcal{F}(t')}{\partial f^\beta(t')} G^{\mathrm{A}}(t',t) \biggr],
\label{eq:response_pm2}
\end{align}\end{widetext}
where the indices for the quantum numbers are omitted and $\Tr$ denotes the trace over the quantum numbers.
The retarded and advanced Green functions in Eq.~\eqref{eq:response_pm2} guarantee the causality: $\chi^{\mathrm{pm}}(t,t')\propto \theta(t-t')$.

We consider the optical conductivity for an example.
The current density and the coupling Hamiltonian between the vector potential and the electrons are given in Eqs.~\eqref{eq:conductivity_current} and \eqref{eq:conductivity_coupling}.
Then, $O_{\mu\nu}^\alpha$ and $\mathcal{F}_{\mu\nu}^\alpha$ in Eqs.~\eqref{eq:response_operator} and \eqref{eq:response_coupling} are identified as
%\begin{subequations}
\begin{align}
O_{\bm{k}s,\bm{k}'s'}^{\alpha}({t}) &= \delta_{\bm{k}\bm{k}'} \delta_{ss'} v_{\bm{k}-\bm{A}({t}),s}^\alpha/N, \\
\mathcal{F}_{\bm{k}s,\bm{k}'s'}({t}) &= -\delta_{\bm{k}\bm{k}'} \delta_{ss'} (\varepsilon_{\bm{k}-\bm{A}({t}),s} - \varepsilon_{\bm{k}s}),
\end{align}%\end{subequations}
respectively, where $\bm{v}_{\bm{k}s}=\partial_{\bm{k}} \varepsilon_{\bm{k}s}$.
The derivatives with respect to the vector potential are given by
%\begin{subequations}
\begin{align}
\frac{\delta O_{\bm{k}s,\bm{k}'s'}^\alpha({t})}{\delta A^{\beta}({t}')}
&= - \delta({t}-{t}') \delta_{\bm{k}\bm{k}'} \delta_{ss'} \frac{1}{N} \frac{\partial^2 \varepsilon_{\bm{k}-\bm{A}(t),s}}{\partial k^\alpha \partial k^\beta}, \\
\frac{\delta \mathcal{F}_{\bm{k}s,\bm{k}'s'}(\bar{{t}})}{\delta A^{\beta}({t}')}
&= \delta(\bar{{t}}-{t}') \delta_{\bm{k}\bm{k}'} \delta_{ss'} v_{\bm{k}-\bm{A}(t'),s}^\beta.
\end{align}%\end{subequations}
By substituting these equations into Eqs.~\eqref{eq:response_dia} and \eqref{eq:response_pm2}, and using relations: $G^{\mathrm{A}}_{\mu\nu}({t}',{t})^* =G_{\nu\mu}^{\mathrm{R}}({t},{t}')$ and $G_{\mu\nu}^{<}({t},{t}')^* = -G_{\nu\mu}^{<}({t}',{t})$, we obtain Eqs.~\eqref{eq:response_dia_result} and \eqref{eq:response_pm_result}.

\bibliography{reference}
\end{document}